\documentclass[lettersize,journal]{IEEEtran}
\usepackage{amsmath,amsfonts}
\usepackage{array}
\usepackage[caption=false,font=normalsize,labelfont=sf,textfont=sf]{subfig}
\usepackage{textcomp}
\usepackage{stfloats}
\usepackage{url}
\usepackage{verbatim}
\usepackage{graphicx}
\def\BibTeX{{\rm B\kern-.05em{\sc i\kern-.025em b}\kern-.08em
    T\kern-.1667em\lower.7ex\hbox{E}\kern-.125emX}}
\usepackage{balance}

\usepackage{amsmath}
\usepackage{amssymb}
\usepackage{tikz-cd}
\usepackage{caption}
\usepackage{graphicx}

\usepackage{hyperref}
\usepackage{tikz,pgfplots}
\usepackage{cite}
\usepackage{amsmath,amssymb,amsfonts}
\usepackage{algorithm}
\usepackage{algpseudocode}
\algnewcommand{\LeftComment}[1]{\Statex \(\triangleright\) #1}
\usepackage{graphicx}
\usepackage{textcomp}
\usepackage{xcolor}
\newcommand{\simu}{\mathcal{S}}

\newcommand{\expect}{\mathsf{Expect}}
\newcommand{\deposit}{\mathsf{Deposit}}
\newcommand{\reward}{\mathsf{Reward}}
\newcommand{\withdraw}{\mathsf{Withdraw}}
\newcommand{\msg}{\mathit{msg}}
\newcommand{\pool}{\mathit{pool}}
\newcommand{\expects}{\mathit{expects}}

\usepackage[font=footnotesize,labelformat=simple]{subcaption}

\definecolor{newblue}{rgb}{0.55,0.66,0.85} 
\definecolor{newred}{rgb}{1.00,0.70,0.70} 
\definecolor{newbrown}{rgb}{0.95,0.69,0.52} 

\usepackage{stfloats, caption}%
\usepackage{ragged2e}
\usepackage{cite}
\usepackage{amsmath,amssymb}
\usepackage{graphicx}
\usepackage{textcomp}
\usepackage{xcolor}

\usepackage{footnote}
\usepackage{tcolorbox}

\usepackage[misc]{ifsym}
\usepackage[justification=centering]{caption}

\usepackage{epsfig,endnotes,amsfonts,bm,tikz,pgfplots}
\usepackage{indentfirst}
\usetikzlibrary{intersections}
\usetikzlibrary{calc}

\usepackage{framed}
\usepackage{enumerate}
\usepackage[
	lambda,
	advantage,
	operators,
	sets,
	adversary,
	landau,
	probability,
	notions,
	ff,
	mm,
	primitives,
	events,
	complexity,
	asymptotics,
	keys
	]{cryptocode}
\createprocedurecommand{functionality}{}{\textbf{Functionality~}}{mode=text}
\usepackage{tcolorbox}

\newtheorem{myDef}{Definition} 

\newtheorem{myLem}{Lemma}
\newtheorem{myThr}{Threat}

\newcommand{\abeglobalsetup}{\textsf{GlobalSetup}}
\newcommand{\abeauthsetup}{\textsf{ABEAuthSetup}}
\newcommand{\abeencrypt}{\textsf{ABEEncrypt}}
\newcommand{\abekeygen}{\textsf{ABEKeyGen}}
\newcommand{\abedecrypt}{\textsf{ABEDecrypt}}

\newcommand{\abekeygenenc}{\textsf{ABEEncKey}}
\newcommand{\getkey}{\textsf{getKey}}
\newcommand{\genProofs}{\textsf{genProofs}}

\newcommand{\ek}{\textit{EK}}
\newcommand{\gid}{\textit{GID}}
\newcommand{\gp}{\textit{GP}}
\newcommand{\acp}{\textit{acp}}

\newcommand{\pku}{pk_u}

\newcommand{\true}{\textsf{true}}

\newcommand{\proofs}{\textsf{proofs}}

\newcommand{\checkkey}{\textsf{checkKey}}

\newcommand{\judgeattrs}{\textsf{judgeAttrs}}
\newcommand{\calc}{\textsf{calc}}

\newcommand{\setup}{\textbf{setup}}
\newcommand{\encrypt}{\textbf{encrypt}}
\newcommand{\vrf}{\textbf{verify}}
\newcommand{\access}{\textbf{access}}
\newcommand{\request}{\textbf{request}}

\newcommand{\kdf}{\textsf{kdf}}
\usepackage{pifont}
\newcommand{\cmark}{\ding{51}}%
\newcommand{\xmark}{\ding{55}}%

\usetikzlibrary{spy,backgrounds}
\usepackage{ragged2e}

\pgfplotsset{colormap={mycolormap}{
    rgb255(0cm)=(0,155,255);
    rgb255(1cm)=(0,155,255);
    rgb255(2cm)=(0,155,255)
}}

\begin{document}


\title{Data sharing in the metaverse with key abuse resistance based on decentralized CP-ABE}

\author{Liang~Zhang, 
        Zhanrong~Ou,
        Changhui~Hu,
        Haibin~Kan$^*$,
        Jiheng~Zhang
\thanks{Liang Zhang, Zhanrong Ou and Changhui Hu are with the School of Cyberspace Security (School of Cryptology), Hainan University, Haikou 570228, China. Liang Zhang is also with Department of Industrial Engineering and Decision Analytics, Hong Kong University of Science and Technology. 
}.
\thanks{Haibin Kan is with Shanghai Key Laboratory of Intelligent Information Processing, School of Computer Science, Fudan University, Shanghai 200433, China, and with Shanghai Engineering Research Center of Blockchain, Shanghai 200433, China. Haibin Kan is also with Zhuhai Fudan Innovation Research Institute, Zhuhai city 518057, China. E-mail: \url{hbkan@fudan.edu.cn}}.
\thanks{Jiheng Zhang is with Department of Industrial Engineering and Decision Analytics, Hong Kong University of Science and Technology.
}.
\thanks{This work was supported National Natural Science Foundation of China (Nos. 62272107 and 62262012), the National Natural Science Foundation of China for Young Scientists (No. 62302129), Hainan Provincial Natural Science Foundation of China (No. 622RC617), Hainan Province Key R\&D plan project (No. ZDYF2024GXJS030) and the HK RGC General Research Fund (Nos. 16208120 and 16214121).}
\thanks{Manuscript received June 19, 2024; revised October 29, 2024 and November 27, 2024; accepted December 3, 2024.}
\thanks{*Corresponding author: Haibin Kan}

}
\IEEEtitleabstractindextext{%
\begin{abstract}
Data sharing is ubiquitous in the metaverse, which adopts blockchain as its foundation. Blockchain is employed because it enables data transparency, achieves tamper resistance, and supports smart contracts. However, securely sharing data based on blockchain necessitates further consideration. Ciphertext-policy attribute-based encryption (CP-ABE) is a promising primitive to provide confidentiality and fine-grained access control. 
Nonetheless, authority accountability and key abuse are critical issues that practical applications must address. Few studies have considered CP-ABE key confidentiality and authority accountability simultaneously. To our knowledge, we are the first to fill this gap by integrating non-interactive zero-knowledge (NIZK) proofs into CP-ABE keys and outsourcing the verification process to a smart contract. To meet the decentralization requirement, we incorporate a decentralized CP-ABE scheme into the proposed data sharing system. Additionally, we provide an implementation based on smart contract to determine whether an access control policy is satisfied by a set of CP-ABE keys. 
We also introduce an open incentive mechanism to encourage honest participation in data sharing.
Hence, the key abuse issue is resolved through the NIZK proof and the incentive mechanism. 
We provide a theoretical analysis and conduct comprehensive experiments to demonstrate the feasibility and efficiency of the data sharing system. Based on the proposed accountable approach, we further illustrate an application in GameFi, where players can play to earn or contribute to an accountable DAO, fostering a thriving metaverse ecosystem.
\justifying
\end{abstract}

}

\maketitle

\IEEEdisplaynontitleabstractindextext

\section{Introduction}
\label{sec:introduction}
The metaverse is going to mix virtual and real worlds more tightly~\cite{Gan2023Web3}. 
In the metaverse, users need to frequently exchange information to enable collaboration across enterprises or among individuals. 
Traditional data sharing approaches~\cite{Dong2014DS,Xiong2018DS,Wang2016DS} rely on a central server or an authority to accomplish data exchange. However, this leads to a single point of failure. It is widely recognized that blockchain is a suitable foundation to build connection between virtual and real worlds. GameFi~\cite{Gadekallu2023Meta} exemplifies a core component of the metaverse, showcasing how blockchain can transform digital experiences. GameFi platforms give players greater control over their assets and data, such as in-game items like weapons and skins. Additionally, GameFi operates within its own economy, allowing users to buy, sell, and trade virtual goods and services. This ecosystem enhances engagement through rewards, challenges, and achievements, making the gaming experience more immersive and rewarding.

Blockchain is seen as a crucial technology to provide a trustworthy approach to store and exchange data. Also, blockchain is the most friendly platform to host smart contract. Previous data sharing systems~\cite{Yu2021DS,Feng2021DS} have shown that blockchain can help enhance convenience and credibility.
In data sharing, blockchain can be adopted as a data registry or data repository~\cite{Zhang2022DS,Wang2020DS}. Besides, blockchain's smart contract execution engine can be regarded as a trustworthy verifier~\cite{Zhang2022DKG}. 
However, there are obstacles when designing data access control on a blockchain due to its transparency, which may bring disastrous risks to data security.
The primary security risks include privacy leakage and unauthorized access.

Ciphertext-policy attribute-based encryption (CP-ABE)~\cite{Bethencourt2007CPABE,Lewko2011DABE,Rouselakis2015EfficientSL} is a promising cryptographic primitive to provide both confidentiality and access control simultaneously. In CP-ABE, the ciphertext is embedded with an access control policy and the decryption key is generated with an attribute set. Only when the set satisfies the policy can the ciphertext be decrypted. Thus, encryptors can protect data privacy using proper access control policy, without caring about the identities of decryptors. That is to say, CP-ABE allows a data owner to share data in a one-to-many paradigm. By incorporating an access control policy in a ciphertext, CP-ABE enables fine-grained access control. In addition, multi-authority CP-ABE~\cite{Rouselakis2015EfficientSL,Lewko2011DABE}, also known as decentralized CP-ABE, allows multiple distributed attribute authorities to generate keys autonomously and jointly in decentralized applications. 


However, CP-ABE keys are sensitive information that should be distributed securely and efficiently to users. 
Key abuse mainly includes illegal authorization and key disclosure~\cite{Hei2021ABE}. 
Illegal authorization means that dishonest authorities may generate keys for some unauthorized data users. 
Key disclosure implies that a dishonest data user may intentionally expose the decryption keys for profit or other illegal usage, leading to risks such as impersonation attacks or unauthorized decryption..
Researches~\cite{Ning2015ACPABE,Yu2017ACPABE,Li2020HCPABE} investigate methods to track the user who leaks the key by mapping the key to the user. However, these schemes require suspected keys to be exposed for verification or audition.

To further uncover misbehavior of authorities, some CP-ABE schemes~\cite{Ning2015ACPABE,Yu2017ACPABE,Li2020HCPABE,Liu2022BACPABE,Yang2022CPABE,Liu2012TCPABE} are dedicated to achieving the property of auditability or accountability. Accountability means that malicious authorities should be caught if they issue illegal keys to authorized users or legitimate keys to unauthorized users. A common approach~\cite{Ning2015ACPABE,Yu2017ACPABE,Li2020HCPABE,Liu2022BACPABE,Yang2022CPABE} is to define an algorithm, which inputs a decryption key and judges whether it is well-formed or finds out its owner. 
However, there are major limitations in practice. On one hand, decryption keys should be exposed if they need to be audited. On the other hand, dishonest users may hide decryption keys as if they have not received the keys. 
Some other works employ blockchain to gain public auditability and use encryption technology to preserve the privacy of CP-ABE keys~\cite{Yuan2017CPABE,Qin2021AC}. Also, an incentive mechanism based on blockchain is adopted to implement key traceability~\cite{Hei2021ABE}. 

In this paper, we envision a data sharing framework for the metaverse with accountable authorities applying a decentralized CP-ABE scheme~\cite{Rouselakis2015EfficientSL}. 
We depict the data sharing paradigm as below. Firstly, a data user, who wants to access an owner's data, deposits some digital currency in a smart contract. Then, attribute authorities generate CP-ABE keys and encrypt them with the user's public key. Also, NIZK proofs are attached. Next, authorities publish the encrypted CP-ABE keys and NIZK proofs on a blockchain, where the encrypted keys are recorded and verified. Meanwhile, the owner and honest authorities will be rewarded by partitioning the deposits. Thereafter, the user obtains a set of valid CP-ABE keys using its private key. Finally, the user can acquire the owner's shared data. 
We further apply the proposed framework in GameFi, enabling players (or providers) to sell (or issue) in-game items with flexible access control. This approach provides a transparent and secure system for players to engage in play-to-earn activities with accountable DAOs.

Our contributions of this paper are summarized as below:
\begin{enumerate}
    \item We propose a secure and efficient blockchain-driven data sharing framework for the metaverse leveraging decentralized CP-ABE, guaranteeing data confidentiality with fine-grained access control.
    \item We are the first to consider CP-ABE key privacy and accountability simultaneously based on blockchain. Non-interactive zero-knowledge (NIZK) proofs are generated to guarantee accountability and they are automatically verified through a smart contract.
    \item We implement the CP-ABE access control policy in smart contract. Further, an open incentive mechanism is proposed to encourage honest behaviors. Based on the mechanism, key abuse problems and collusion attacks are eliminated.
    \item We give concrete security analysis of the proposed method and conduct comprehensive experiments on the Ethereum. The experimental results show the feasibility and practicality of our proposed scheme. 
\end{enumerate}






\section{Preliminaries}

\subsection{Bilinear mapping}
\label{bilinearmaps}
$\GG_1$, $\GG_2$ and $\GG_T$ are groups of prime order $p$. Let $g_1$, $g_2$ be a generator of $\GG_1$ and $\GG_2$ repectively and $e: \GG_1$ $\times$ $\GG_2$ $\to$ $\GG_T$ be a bilinear map with following properties:
\begin{enumerate}
    \item  Bilinearity: for all $\mu$ $\in$  $\GG_1$, $\upsilon$ $\in$  $\GG_2$ and $a,b \in$ $\ZZ_p$, we have $e(\mu$$^a$, $\upsilon$$^b$) = $e(\mu$, $\upsilon$)$^{ab}$.
    \item Nondegeneracy: $e(g_1, g_2) \ne$ 1.
\end{enumerate}
Usually, the operation $e$ is efficiently computable. The resulting group $\GG_T$ is called a bilinear mapping group.

\subsection{Linear relationship Sigma protocol and NIZK proof}
\label{sec:linear_sigma_protocol}
The Sigma protocol~\cite{Boneh2023} is a three-move protocol between a prover and a verifier that can be used to prove knowledge of certain relationships without revealing any information about the knowledge itself.
In linear relationship Sigma protocol, a prover P can prove zero knowledge of witness $X=\{x_1,...,x_n\}$ for statement $Y$ as below, where $Y=h_1^{x_1}...h_n^{x_n}$ and $h_1,...,h_n$ are generators of $\GG$.

\begin{center}
\fbox{
\scalebox{0.999}{
\pseudocode{%
    {\rm P}(X,Y) \< \< {\rm V} \\[][\hline]
    x^\prime_1,...,x^\prime_n \xleftarrow[]{R} \ZZ_p \< \< \\
    \left\{
        \begin{array}{l}
          b=g_1^{x^\prime_1}...g_n^{x^\prime_n} \tabularnewline 
          c=Hash(Y, b) \tabularnewline 
          w_1=x_1^\prime+c\cdot x_1 \tabularnewline 
          ...\tabularnewline
          w_n=x_n^\prime+c\cdot x_n
      \end{array} \right. \<\hspace{-0.3cm} \sendmessageright{length=2.cm, top={$y,b, c$},bottom={$ \{w_1,...,w_n\}$}} \<\hspace{-0.3cm} g_1^{w_1}...g_n^{w_n} \overset{\text{?}}{=}  b\cdot Y^c
}
}
}
\end{center}
$Hash$ is modeled as a random oracle, as required by Fiat-Shamir heuristic~\cite{FS1986transform}. Denote the conversation $(b, c, \{w_1,...,w_n\})$ as $\proofs$ for the prover P. It can be concluded that above protocol constructs non-interactive zero knowledge (NIZK) system between P and V. The required security properties of Sigma protocol are given by Definition~\ref{def:sigma_properties} in the Appendix~\ref{sec:security_properties}.

\subsection{Decentralized CP-ABE}
Decentralized CP-ABE scheme~\cite{Rouselakis2015EfficientSL,Lewko2011DABE} contains following five algorithms:
\begin{itemize}
    \item $\gp \gets \abeglobalsetup(\Lambda)$. It takes in the security parameter $\Lambda$ and outputs global parameters $\gp$. 
    \item $\sk_\theta, \pk_\theta \gets \abeauthsetup(\gp, \theta)$. Each authority $\theta$ takes $\gp$ as input to produce a key pair ($\sk_\theta$, $\pk_\theta$). 
    \item $C \gets \abeencrypt(M, acp, \gp, \{\pk_\theta\})$. The algorithm takes in $\gp$, a message $M\in \GG_T$, an access control policy $acp$ and a set of public keys. It outputs a ciphertext $C$.
    \item $K_0,K_1 \gets \abekeygen(\gid, \gp, u, \sk_\theta)$. The algorithm takes in an identity $\gid$, $\gp$, an attribute $u$ belonging to the authority $\theta$, and an authority's secret key $\sk_\theta$. It produces a decryption key $K_{\gid,u}$. For convenience, we use $K_0, K_1$ and omit $\gid$ and $u$ where RW CP-ABE is leveraged in this paper.
    \item $M \gets \abedecrypt(\gp, C, \{K_{\gid,u}\})$. The decryption algorithm takes in $\gp$, the ciphertext $C$, and a collection of decryption keys. Only if the collection of decryption keys satisfies the access control policy in the ciphertext, it outputs the message $M$.
\end{itemize}

We concentrate on the authority accountability feature. $\abekeygen$ is the interface to issue keys for decryptors. Our idea of accountability can be applied to most CP-ABE schemes, in which keys are group elements (i.e., public key-style). Without loss of generality, we choose Rouselakis and Waters's (RW)~\cite{Rouselakis2015EfficientSL} decentralized CP-ABE scheme.  Figure~\ref{fig:rwkeygen} depicts the $\abekeygen$ RW CP-ABE algorithm.
\begin{figure}[!hpbt]
\centering
\fbox{
    \begin{minipage}{0.93\linewidth}
\underline{$\abekeygen(\gid, \gp, u, \sk_\theta)\rightarrow (K_0, {K_1}):$} \\
         \hspace*{0.5cm}  $d_\theta \xleftarrow[]{R} \ZZ_p$\\
         \hspace*{0.5cm}  $\left\{
          \begin{array}{l}
              {K_0} = g_1^{\alpha_\theta} H(\gid)^{\beta_\theta}F(u)^{d_\theta}\tabularnewline
              {K_1} = g_2^{d_\theta}
          \end{array} \right.$\\
\end{minipage}
}
\caption{\label{fig:rwkeygen}RW CP-ABE $\abekeygen$ algorithm}
\end{figure}

where $\sk_\theta=(\alpha_\theta,\beta_\theta)$ and $\pk_\theta=(g_1^{\alpha_\theta}, g_2^{\alpha_\theta}$, $g_2^{\beta_\theta}$, $e(g_1,g_2)^{\alpha_\theta})$ are authority $\theta$'s key pair; $g_1$ and $g_2$ are generators of group $\GG_1$ and $\GG_2$, respectively; $H$ maps global identity $\gid$ to an element on $\GG_1$; $F$ maps an arbitrary string to an element on $\GG_1$. Note that $g_1^{\alpha_\theta}$ is exposed in our application. It does not reduce the scheme security, since $K_0$ is also protected by $\beta_\theta$ and a blinded factor $d_\theta$.

\subsection{Ethereum blockchain}
Ethereum~\cite{Wood2014Eth} is one of the most innovative blockchain. Its security is based on distributed storage using a cryptographic chain of blocks and the consensus algorithm. Smart contract is a program that aims at building agreement among distributed nodes. 
Once the contract is executed in Ethereum virtual machine (EVM), it is written onto the blockchain, making it traceable and irreversible. Each smart contract has a unique address, which is created when the user deploys the contract. 
Pairing-check is a cryptographic operation that judges whether multiple group elements satisfy specific group mapping defined in EIP-197~\cite{EIP197}. 
However, pairing-check only outputs a boolean value, rather than providing a group mapping (or bilinear mapping) result.
The gas system is designed to eliminate distributed denial-of-service (DDoS) attacks on the Ethereum network. Every computation and storage costs a certain amount of gas (digital currency) from the transaction invoker. The used gas is packed as a reward for miners. In Ethereum, Ether is widely adopted digital currency units, and 1Ether=$10^{18}$wei.


\section{System model and threat model}
\subsection{System entities}
\begin{enumerate}
    \item \textbf{Data owner}: A data owner is the possessor of data (i.e., digital assets, such as communication keys, NFTs, files and other electronic information). Data owners share data, which is encrypted with RW CP-ABE $\abeencrypt$ algorithm, for digital currency as benefits. 
    \item \textbf{Data user}: A data user with ($pk_u, sk_u$) is the demanding side who wants to use data. Data users pay for the data when they are authorized to access it (i.e., they are able to decrypt the corresponding RW CP-ABE ciphertexts).
    \item \textbf{Attribute Authorities (AAs)}: Each authority $\theta$ with ($\pk_\theta, \sk_\theta$) earns digital currency by offering key issuance services for data users. Authorities are independently and autonomously distributed all over the world. 
    \item \textbf{Blockchain}: Blockchain is used as a decentralized, available and tamper-resistant service, providing reliable storage and verifiable computation. In our proposed system, it is used to manage (i.e., store and verify) RW CP-ABE keys. Particularly, we choose Ethereum, a permissionless ecosystem, as the underlying blockchain.
    \item \textbf{Data storage provider (DSP)}: Data storage provider is the service to store CP-ABE ciphertext from data owners. 
\end{enumerate}


\subsection{System assumption and threats}
We aim at achieving a blockchain and CP-ABE based privacy-preserving data sharing framework in the metaverse. Specifically, CP-ABE keys are managed transparently yet accountably on the blockchain and key abuse resistance should be considered.
Global identifier $\gid$ is unique to each sharing instance. An attribute string $u$ can be used in different sharing instances.

If an entity abides by the proposed scheme, it is assumed to be honest. Otherwise, it is regarded to be dishonest or malicious. We assume an $\acp$ in a sharing instance cannot be satisfied by keys issued only by malicious authorities. Hence, we do not consider the collusion attack between user and authorities.
Authorities are required to stake digital currency to offer key issuance services. If they are found to act dishonestly, their staked digital currency will be forfeited. The DSP is honest but curious, and its benefits within the system are not specified. The authorities are authenticated to \textbf{DSP}.
The potential threats in our framework include:
\begin{myThr}\label{thr:1} Adversaries may try to construct/infer a key.\end{myThr}

\begin{myThr}\label{thr:2} A malicious user may forge a key for an attribute $u'$ which he/she does not have.\end{myThr}

\begin{myThr}\label{thr:3} Malicious authority may generate invalid keys or fake proofs in key issuance.\end{myThr}

\begin{myThr}\label{thr:4} Malicious users may try to obtain owner's ciphertext or digital assets by impersonation attack.\end{myThr}

\begin{myThr}\label{thr:5} Malicious users may collude to obtain and decrypt a data owner's ciphertext or digital assets.\end{myThr}

\subsection{System goals}
\label{sec:goals}
The goals of the system include the following properties.
\begin{enumerate}
    \item \textit{Tamper-resistance}: Data authenticity, validity, and permanence are ensured when data is stored on the chain. 
    \item \textit{Decentralization}: The system is realized without a trusted third party, avoiding the single point of failure problem.
    \item \textit{Key privacy}: CP-ABE keys are encrypted and delivered through blockchain against eavesdropping attack.
    \item \textit{Data privacy}: Data is confidential to unauthorized users.
    \item \textit{Forward-secrecy}: If the CP-ABE keys of some sharing instance ($\gid$) are compromised, the privacy of previously shared data is still guaranteed.
    \item \textit{Incentives}: Data users and authorities are rewarded for participating in the data sharing scheme.
    \item \textit{Accountability}: An encrypted key can be checked whether it is correct for the designated user with $pk_u$.
    \item \textit{Key abuse resistance}: Illegal key authorization is eliminated and key disclosure are useless for the system.
\end{enumerate}


\section{The proposed system}
\subsection{High-level overview}
\begin{figure}[!h]
    \centering
    \scalebox{0.13}{
    \includegraphics{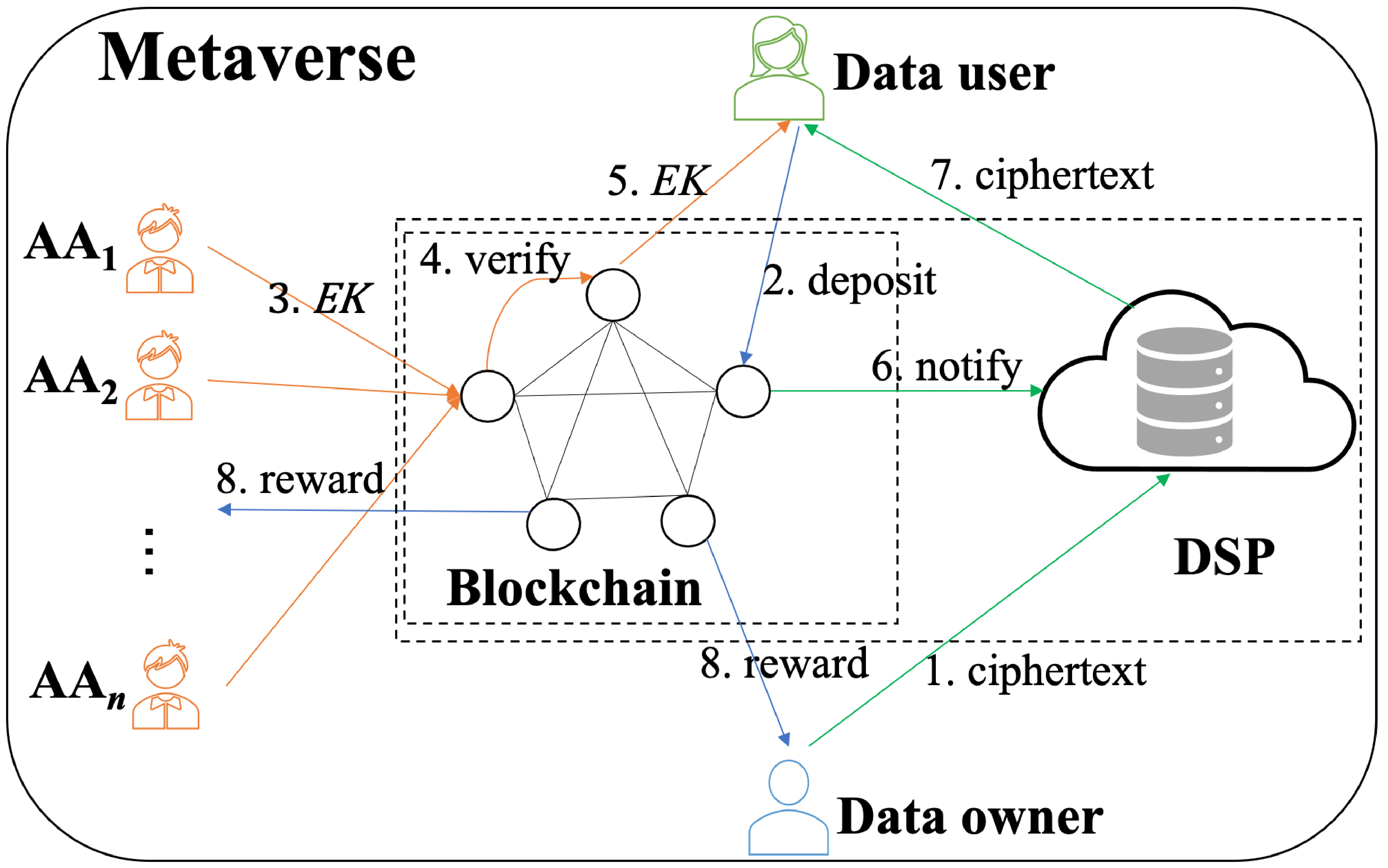}
    }
    \caption{High-level overview of the system architecture}
    \label{fig:arch}
\end{figure}

In the metaverse, data sharing requires no central trusted third party (TTP). Hence, blockchain is leveraged to play the role of trusted proxy and its smart contract enables parties in the metaverse to trade digital assets fairly. Moreover, we incorporate a multi-authority CP-ABE algorithm to decentralize the identities of the players or creators.
Figure~\ref{fig:arch} depicts the high-level overview of the system architecture. 

Specifically, \textbf{DSP} is used to store ciphertexts, where symmetric encryption is used to wrap the CP-ABE scheme to encrypt arbitrary data. Blockchain provides tamper-proof storage and trustworthy verification computation. Attribute authorities generate CP-ABE keys and encrypt them with data user's public key. Then, they publish the encrypted CP-ABE keys to the blockchain, which records and verifies each of the encrypted keys. Finally, the data user gets valid encrypted CP-ABE keys from the blockchain and ciphertext from the \textbf{DSP} to obtain the data owner's shared data.
The green arrows show the actions related to data users' ciphertext; The blue arrows demonstrate how the incentives (digital currency) flow; The orange arrows introduce how \textbf{AA}'s encrypted keys are transferred and verified.

\subsection{Accountability of authorities via smart contract}
\label{sec:accountability}
In CP-ABE-based applications, authorities have to use a private secure channel to deliver decryption keys after invoking the $\abekeygen$ algorithm. To allow authorities to incorporate public channels, we slightly modify the $\abekeygen$ algorithm to the $\abekeygenenc$ algorithm. Particularly, by replacing $g_1$ in $K_0$ with the targeted user's public key $\pku=g_1^y$, we obtain $\ek_0$. Hence, the encrypted keys are ($\ek_0, \ek_1$) and they can be delivered through blockchain.

\underline{$\abekeygenenc(\gid, \gp, u, \sk_\theta, \pku)\rightarrow (\ek_0, \ek_1):$} 
\begin{align}
 & d_\theta \xleftarrow[]{R} \ZZ_p\notag\\
 & \left\{
  \begin{array}{l}
      {\ek_0} = (\pku)^{\alpha_\theta} H(\gid)^{\beta_\theta}F(u)^{d_\theta}\tabularnewline
      {\ek_1} = g_2^{d_\theta}\tabularnewline
  \end{array} \right.\notag
\end{align}

Obviously, the encrypted key ($\ek_0,\ek_1$) cannot be directly used to decrypt any RW CP-ABE ciphertext.

\begin{myLem}\label{lem:enckey_secure}    
The $\abekeygenenc$ algorithm is resistant to eavesdropping attacks, i.e., no adversary can recover a decryption key from eavesdropping the blockchain.
\end{myLem}
\textbf{Proof:} Obviously, an adversary $\adv$ cannot learn useful information from $\ek_1$. Neither, $\adv$ learns nothing from $\ek_0$ which can be regarded as a Pederson commitment~\cite{Pedersen1991VSS} of the authority $\theta$. Due to the perfect hiding of Pederson commitment, $\adv$ cannot forge a valid commitment/ciphertext $\ek_0$, even if the secret ($\alpha_\theta, \beta_\theta$) are used in multiple commitments/ciphertexts, as long as $log_{pk_u}F(u)$ and $log_{H(\gid)}F(u)$ are unknown. Actually, $\ek_0$ hides $K_0$ using DH exchange symmetric key $(g_1^{\alpha})^{y-1}$. 
Note that $K_0$ can be treated as another Pederson commitment with both $g_1$ and $H(\gid)$ as bases, without exposing the secret ($\alpha_\theta, \beta_\theta$) of the authority $\theta$.
Therefore, $\adv$ cannot obtain a key given all encrypted keys are stored on publicly on blockchain, addressing Threat~\ref{thr:1}.

By proving Lemma~\ref{lem:enckey_secure}, we know that only the targeted user, with secret key $sk_u=y$, can recover a valid decryption key as the $\getkey$ algorithm shows.

\underline{$\getkey(\ek_0, \ek_1, g_1^{\alpha_\theta}, y)\rightarrow (K_0, K_1):$}
\begin{align}
&\hspace{1cm}\begin{array}{l}
    K_0=\ek_0/{(g_1^{\alpha_\theta})}^{y-1}=g_1^{\alpha_\theta}H(\gid)^{\beta_\theta}F(u)^{d_\theta} \tabularnewline
    K_1=\ek_1=g_2^{d_\theta}
\end{array}\notag
\end{align}

\begin{myLem}\label{lem:2}
Malicious users cannot forge a key for an attribute
$u'$ which he/she does not have, addressing Threat~\ref{thr:2}.
\end{myLem}
\textbf{Proof:} A user who decrypts ($EK_0, EK_1$) and obtains key ($K_0, K_1$), may forge key via malleability\footnote{This property sometimes is used for achieving the functionality of key delegation~\cite{Bethencourt2007CPABE}.} by calculating $( K_0\cdot F(u)^r, K_1\cdot g_2^r)$, where $r\xleftarrow{R}\ZZ_p$. Obviously, the new key is still bound to the attribute $u$. Hence, the user cannot successfully forge a meaningful key for attribute $u'\neq u$.

Whether an authority has correctly perform the encryption with the $\abekeygenenc$ algorithm needs to be considered, especially when authorities are beneficiaries in the metaverse applications. Therefore, we take advantage of the linear relationship of the Sigma protocol (defined in Section~\ref{sec:linear_sigma_protocol}) to make the encrypted key $(\ek_0, \ek_1)$ publicly verifiable. Hence, the NIZK $\proofs$ are generated as $\genProofs$ depicts.

\underline{$\genProofs(\gid, u, \pku)\rightarrow \proofs:$} 
\begin{align}
&\hspace{1cm}\alpha_\theta^\prime,\beta_\theta^\prime,d_\theta^\prime \xleftarrow[]{R} \ZZ_p\notag\\
&\hspace{1cm}\proofs=\left\{\begin{array}{l}
    \ek_0^\prime=(\pku)^{\alpha_\theta^\prime} H(\gid)^{\beta_\theta^\prime}F(u)^{d_\theta^\prime}\tabularnewline
    \ek_1^\prime=g_2^{d_\theta^\prime}\tabularnewline
    c=Hash(\ek_0, \ek_0^\prime)\tabularnewline
    w_1=\alpha_\theta^\prime + c\cdot \alpha_\theta \tabularnewline
    w_2=\beta_\theta^\prime + c\cdot \beta_\theta \tabularnewline
    w_3=d_\theta^\prime + c\cdot d_\theta
\end{array}\right.
\end{align}

When ($\ek_0, \ek_1, \proofs$) is submitted to a verifier, he/she can check whether $(\ek_0, \ek_1)$ corresponds to a valid decryption key using $\checkkey$. The $\checkkey$ algorithm inputs the sharing instance $\gid$, the attribute $u$, encrypted key ($\ek_0, \ek_1$) and the NIZK $\proofs$, and it is formally defined as below:
\begin{align}
&\hspace{-0.5cm}\underline{\checkkey(\ek_0,\ek_1,\proofs,\gid,u):}\notag\\ 
&(\pku)^{w_1} H(\gid)^{w_2}F(u)^{w_3}\overset{\text{?}}{=}\ek_0^\prime\cdot \ek_0^c\label{eq:checkkey1}\\
&g_2^{w_3}\overset{\text{?}}{=}\ek_1^\prime\cdot \ek_1^c\label{eq:checkkey2}\\
&e(\pku, g_2^{\alpha_\theta})e(H(\gid),g_2^{\beta_\theta})e(F(u),\ek_1) \overset{\text{?}}{=}e(\ek_0,g_2) \label{eq:checkkey3}
\end{align}

\begin{myLem}\label{lem:3}
Malicious authority cannot generate invalid keys or proofs, addressing Threat~\ref{thr:3}.
\end{myLem}
\textbf{Proof:}
The NIZK $\proofs$ in the $\checkkey$ algorithm guarantees honest issuance of $(\ek_0, \ek_1)$ by the authority. The equation~\eqref{eq:checkkey1} is the linear Sigma protocol, proving knowledge of the tuple $\{\alpha_\theta, \beta_\theta, d_\theta\}$. The equation~\eqref{eq:checkkey2} is the linear Sigma protocol, proving knowledge of the $\{d_\theta\}$. Since both $\ek_0$ and $\ek_1$ have leveraged the same $d_\theta$, we adopt the equation~\eqref{eq:checkkey3} to prove the binding relationship of $d_\theta$.
These equations are automatically executed on smart contract.
Moreover, if any equation in $\checkkey$ does not hold, the authority is deemed dishonest, his/her stake will be forfeited as a result. The required security properties of the NIZK proof system is defined in Appendix~\ref{sec:security_properties} and the corresponding security proofs are demonstrated in the Appendix~\ref{sec:security_proof}.


\subsection{Access control in smart contract}
\label{sec:acp_in_sc}

\textbf{DSP}, who manages CP-ABE ciphertext, should be able to judge whether a user is genuinely granted to access the ciphertext. This is done by letting \textbf{DSP} check whether a user indeed has a legal key from authorities. A legal key is a combination of sub-keys from distributed authorities. However, $\checkkey$ defined in Section~\ref{sec:accountability} can only be used to verify each sub-key. Hence, we design a mechanism to judge the validity of a combined key in a smart contract.
Suppose the combined key corresponds to the data user's attribute set $[u]$. In case a malicious user forges an attribute set, $[u]$ only contains the attribute whose corresponding sub-key is verified using $\checkkey$. Thus, only if the data user's attribute set $[u]$ satisfies the $\acp$, the \textbf{DSP} sends the ciphertext to the user. We omit the details on how the user (with $sk_u$) prove himself/herself to \textbf{DSP}.

\begin{algorithm}[h]
\caption{Access control in smart contract}\label{alg:judgeAttrs}
\begin{algorithmic}
\LeftComment{\textcolor{gray}{Global variables}}
\State global $ops \gets []$
\State global $result \gets []$
\end{algorithmic}
\textbf{function} $\calc()$:
\begin{algorithmic}[1]
\LeftComment{\textcolor{gray}{Pop two values from $result$ and push the new result}}
\If{$result$.length $\textless$ 2} return
\EndIf
\State $op=ops$.pop()
\State $t1$ = $result$.pop()
\State $t2$ = $result$.pop()
\If{$op$ == ``AND"}       
\State $result$.push($t1\&t2$)
\ElsIf{$op$ == ``OR"}
\State $result$.push($t1\|t2$)
\EndIf
\end{algorithmic}
\textbf{function} $\judgeattrs([u],\acp)$:
\begin{algorithmic}[1]
\LeftComment{\textcolor{gray}{Scan $acp$ and invoke $\calc$ to obtain the final result}}
\State $words \gets split(\acp, ``\ ")$
\While{$word \in words$}
    \If{$word$ == ``AND" $\|$ $word$ == ``OR"}
        \If{$ops$.length $\textgreater$ 0 $\&$ $ops$.top() != $``("$}
        \State \calc()
        \EndIf
        \State $ops$.push(word)
    \ElsIf{$word$ == ``("}
        \State $ops$.push(``(")
    \ElsIf{$word$ == ``)"}
        \State $top\gets ops.$top()
        \While{$top$ != ``)"}
            \State $top=ops.$pop()
            \State \calc()
        \EndWhile
    \Else{}
        \State $result$.push($[u]$.contains(word)==\true)
    \EndIf
\EndWhile
\While{$ops$.length $\textgreater$ 0}
    \State \calc()
\EndWhile
\State \textbf{Return} result[0]
\end{algorithmic}
\end{algorithm}
The $\judgeattrs$ function in Algorithm~\ref{alg:judgeAttrs} demonstrates how to judge whether an attribute set $[u]$ satisfies the access control policy $\acp$. Access control policy $\acp$ can be regarded as propositions, only given a proper condition (i.e., the attribute set $[u]$) can they be determined to be $\true$. 
The $\judgeattrs$ function calculates $\acp$ as an infix expression, given values $[u]$. Particularly, we use two stacks $ops$ and $result$ to store operations (i.e., AND/OR) when scanning $\acp$ and values after calculating the operations. $\judgeattrs$ scans the $\acp$ and invokes $\calc$ to calculate the result step by step. $\calc$ is a function to get the two top values from $result$ and it computes a new result according to the last operation from $ops$.

\begin{myLem}\label{lem:4}
Malicious users cannot obtain owner's ciphertext by impersonation attack, addressing Threat~\ref{thr:4}.
\end{myLem}
\textbf{Proof:} \textbf{DSP} sends ciphertext to a user only if attributes in $[u]$ are truly used by authorities and $\judgeattrs$ outputs $\true$. Moreover, by Lemma~\ref{lem:2}, we have proved that a malicious user cannot obtain additional keys regarding the attributes he/she does not have. Hence, malicious users cannot deceive \textbf{DSP} to obtains ciphertext via impersonation attack.

\begin{myLem}\label{lem:5}
Malicious users cannot collude to obtain the data owner's ciphertext, addressing Threat~\ref{thr:5}.
\end{myLem}
\textbf{Proof:} On the one hand, users from different sharing instances cannot combine their keys, due to the use of different $\gid$s. On the other hand, the real attribute set $[u]$ is recorded in smart contract when authorities issue the corresponding encrypted keys. Then, the execution of $\judgeattrs([u],\acp)$ guarantees the authenticity of each unique user, eliminating the possibility of key combination from different users. Therefore, users cannot collude to obtain data owner's ciphertext.

\subsection{Incentive mechanism}
We have shown how to check the honesty of an attribute authority using $\checkkey$ in Section~\ref{sec:accountability}. However, an authority may behave slackly or just abort the system due to no benefits. In this section, we introduce an incentive mechanism to enhance the participation of authorities. In our system, data owners and attribute authorities would gain benefits from data users, due to sharing data and issuing decryption keys, respectively. Furthermore, fairness is guaranteed when owners share data and users make deposits.

\begin{algorithm}[!h]
\caption{Incentive contract}\label{alg:incentive}
\begin{algorithmic}
\LeftComment{\textcolor{gray}{Global variables}}
\State global $\pool \gets \{\}$
\State global $\expects \gets \{\}$
\end{algorithmic}

    \textbf{function} $\expect$($\gid$, ownerVal):\textcolor{gray}{\% by a data owner}
    \begin{algorithmic}[1]
        \State $\expects$[$\gid$] = ownerVal
        \State \textbf{return} True
    \end{algorithmic}
    \textbf{function} $\deposit$($\gid$):\textcolor{gray}{\% by a data user}
    \begin{algorithmic}[1]
        \State $\pool$[$\msg.sender$][$\gid$] = $\msg.value$
        \State \textbf{return} True
    \end{algorithmic}
    \textbf{function} $\withdraw$($\gid$):\textcolor{gray}{\% by a data owner}
    \begin{algorithmic}[1]
    	\State assert($\pool$[$\msg.sender$][$\gid$]$\textgreater$0)
        \State $\msg.sender$.transfer($\pool$[$\msg.sender$][$\gid$])
        \State $\pool$[$\msg.sender$][$\gid$] = 0
        \State \textbf{return} True
    \end{algorithmic}
    \textbf{function} $\reward$($addr_u, addr_o, Addrs,\gid$): \textcolor{gray}{\% by anyone}
    \begin{algorithmic}[1]
        \State assert($\pool$[$addr_u$][$\gid$]$\textgreater expects[\gid]$)
        \State ownerVal=$\expects[\gid]$
        \State $addr_o$.transfer(ownerVal) 
        \State avg=(deposits[$addr_u$][$\gid$]-ownerVal)/$Addrs$.length
        \While{$addr_{aa} \in Addrs$}
            \State $addr_{aa}$.transfer(avg) 
        \EndWhile
        \State deposits[$addr_u$][$\gid$] = 0
        \State \textbf{return} True
    \end{algorithmic}
\end{algorithm}

We introduce an owner-driven approach, in which data owners set a reasonable benefit for attribute authorities when sharing data. The pseudocodes in Algorithm~\ref{alg:incentive} describe the operations each entity performs in the incentive smart contract of the data sharing paradigm. $\expect$ is the function to input an expected profit for a data owner. $\deposit$ is the function to apply for a mortgage for a potential data user. $\withdraw$ is the function to withdraw deposited digital currency for data users. $\reward$ is the function invoked automatically by the smart contract if data sharing is accomplished between data users and data owners.

\begin{figure*}[!t]
    \centering
    \scalebox{0.3}{
    \includegraphics{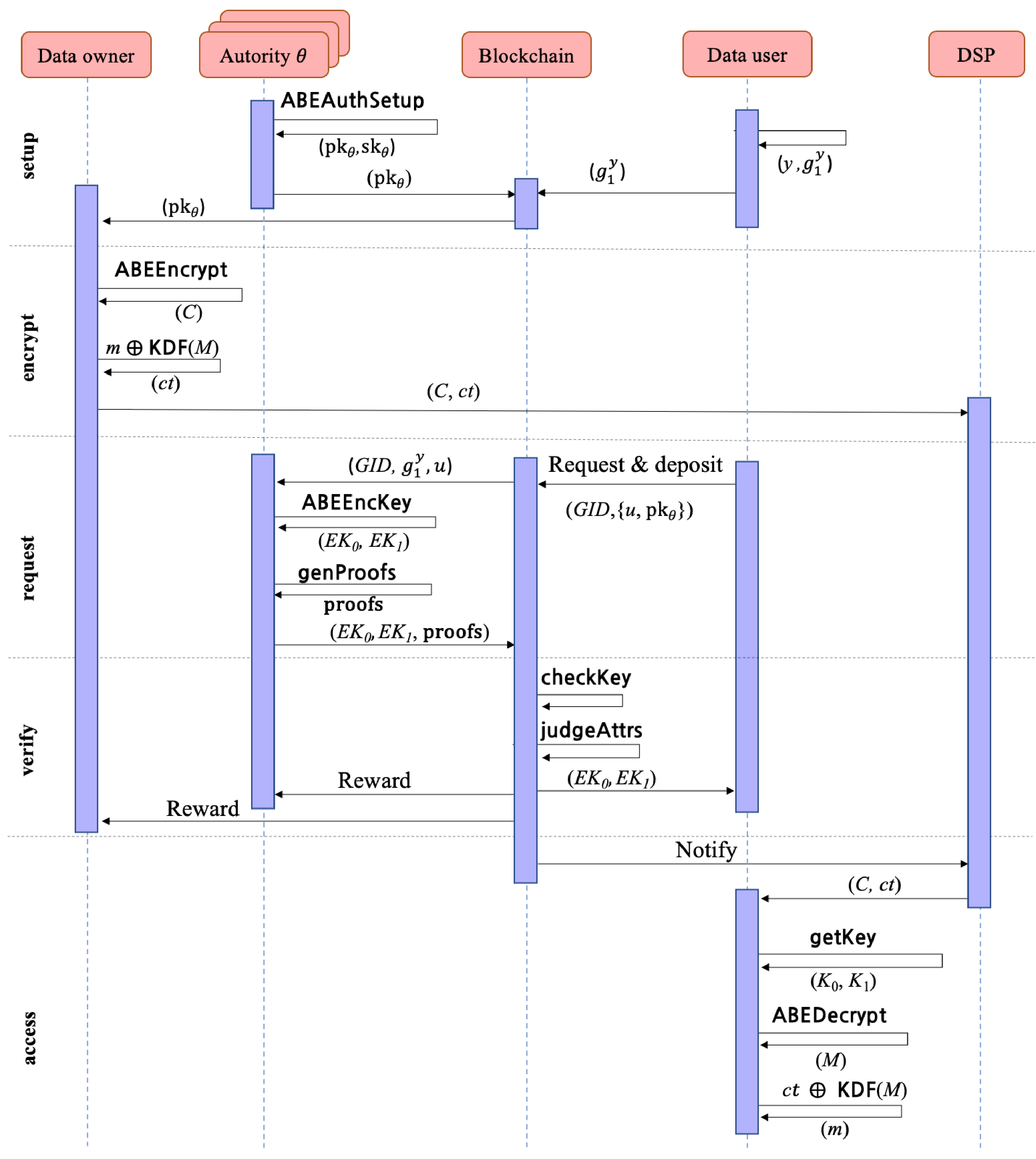}
    }
    \caption{Sequence diagram of the blockchain-driven data sharing paradigm for the metaverse based on CP-ABE}
    \label{fig:dataflow}
\end{figure*}

In the pseudocodes, $msg$ is an embedded variable, where $msg.sender$ represents a transaction invoker; $msg.value$ is the transferred value measured in Ether. $\pool$ and $\expects$ are self-defined global variables to store deposited values and expected profits. $addr_u$, $addr_o$ and $Addrs$ are the addresses of the data user, data owner and honest authorities, respectively.

Also, we can modify the incentive smart contract into a user-driven approach, in which data users are the initiators to put forward a deposit proposal. We omit the details on the modification due to space limitations.

\subsection{Our construction}

Suppose an owner shares its data with users, leveraging the RW decentralized CP-ABE to protect the privacy of the data. We assume the RW CP-ABE $\mathit{GlobleSetup}$ is initialized in a decentralized way. Figure~\ref{fig:dataflow} shows the sequence diagram of the process, which is divided into five phases: $\setup$, $\encrypt$, $\request$, $\vrf$ and $\access$. In our construction, we use $\gid$ as the unique identifier for each data-sharing process.

\begin{enumerate}
    \item $\setup.$ Each authority invokes $\abeauthsetup$ to obtain his/her key pair ($\pk_\theta,\sk_\theta$) and a data user generates his/her key pair ($y, \pku=g_1^y$). Then, public keys are published on a blockchain so that they are shared. Finally, the data owner collects authorities' public keys $\{\pk_\theta\}$.
    \item $\encrypt.$ The data owner encrypts a random value $M$ to get ciphertext $C\gets \abeencrypt(M, acp, \gp, \{\pk_\theta\})$, where $acp$ is a well-designed access control policy. Further, the owner encrypts the data $m$ as $m \oplus \kdf(M)$, where $\kdf$ is a key derivation function and $\oplus$ means exclusive-or operation. Then, the owner stores $(ct, C)$ on DSP. Simultaneously, the data owner sets an expected profit value $v$ by invoking $\expect(\gid, v)$.
    \item $\request.$ The data user sends a request, containing a set of tuples $(\gid, \{u, \pk_\theta\}_{\theta\ in\ \acp})$, to the blockchain after invoking $\deposit(\gid)$ to deposit some digital currency. Then, the blockchain notifies authorities that appeared in $\acp$. Authority $\theta$ fetches the tuple $(\gid, \pku, u)$, where $\pku$ is the user's public key. Next, the authority $\theta$ invokes $\abekeygenenc(\gid,\gp,u,sk_\theta, \pku)$ to produce an encrypted CP-ABE key ($\ek_0, \ek_1$). Simultaneously, the corresponding NIZK $\proofs$ are generated using algorithm $\genProofs$. Finally, the authority $\theta$ sends $(\ek_0, \ek_1, \proofs)$ back to the blockchain.
    \item $\vrf.$ Upon receiving the tuple $(\ek_0, \ek_1$, $\proofs)$ from authority $\theta$, the blockchain smart contract invokes $\checkkey(\ek_0, \ek_1, \proofs, \gid, u)$ algorithm automatically. If $\checkkey$ outputs $\true$, meaning the authority $\theta$ is honest, the blockchain notifies the data user to fetch $(\ek_0, \ek_1)$. To eliminate the off-chain colluding attack by the authorities, the smart contract invokes $\judgeattrs([u],\acp)$ to judge whether $\acp$ in the ciphertext $C$ is satisfied by enough CP-ABE keys $\{\ek_0, \ek_1\}$. Once satisfied, the deposited digital currency is automatically transferred to each authority and the data owner as incentives using $\reward$($addr_u$, $addr_o$, $Addrs$, $\gid$).
    \item $\access.$ Furthermore, the smart contract notifies \textbf{DSP} that the data user is qualified to access the ciphertext $(C, ct)$. After obtaining the ciphertext $(C, ct)$, the data user calculates RW CP-ABE decryption key ($K_0, K_1$) by invoking $\getkey(\ek_0,\ek_1,g_1^{\alpha_\theta},y)$. Next, the data user invokes $\abedecrypt(\gp,C,\{K_0,K_1\})$ algorithm to obtain the value $M$. Finally, the original data $m$ shared by the data user is accessed as $ct \oplus \kdf(M)$.
\end{enumerate}
\textbf{Remark:} If not enough authorities issue decryption keys for a data user, the user can invoke $\withdraw(\gid)$ to withdraw its deposited Ether.

\section{Evaluation}

\subsection{Experiments}
We simulate the proposed system on Ubuntu 22.04-VMware workstation, powered by an Intel(R) Core(TM) i7-9750H CPU @ 2.60GHz and 4GB of RAM. Ganache 2.7.0 is leveraged as the Ethereum test network. The embedded elliptic curve on Ethereum is ``bn128''~\cite{EIP196} and it only supports pairing-check~\cite{EIP197}, rather than bilinear mapping calculation. The onchain codes are written using Solidity language and the offchain codes are developed using Golang. The reason to choose Golang as offchain language lies in Golang is the programming language of EVM. Moreover, Golang has high performance in elliptic curve operations. Table~\ref{table:group_op_cost} has shown that performance of Golang is hundreds of times better than Python, benchmarking based on Ethereum official ``bn128'' library.
The proof-of-concept implementation of the proposed system is published on GitHub.\footnote{\url{https://github.com/AppCrypto/BASICS/}}

\renewcommand{\arraystretch}{1.22}
\begin{table}[h]
    \caption{\centering Comparison of group computation cost (ms)}\label{table:group_op_cost}
    \centering
    \scalebox{1}{
    \begin{tabular}{ |c|c|c|c|c| }
        \hline 
         Language & Multiply$_{\GG_1}$ &Multiply$_{\GG_2}$ & Multiply$_{\GG_T}$ & Pairing \\\hline 
        Python & {10} & {40} & {190} & {700} 
        \\
        Golang & {0.1} & {0.3} & {0.7} & {2} 
        \\\hline 
    \end{tabular}	
    }
\end{table}

\begin{figure*}[!b]
\begin{minipage}[t]{0.24\textwidth}
\centering
\scalebox{0.5}{
\includegraphics{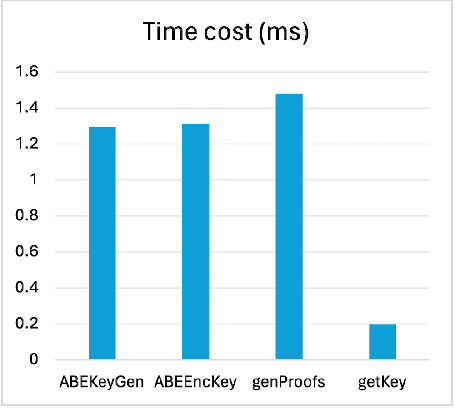}
}
\caption{\label{fig:key_time_cost}Cost of offchain operations (ms)}
\end{minipage}
\begin{minipage}[t]{0.24\textwidth}
\centering
\scalebox{0.5}{
\includegraphics{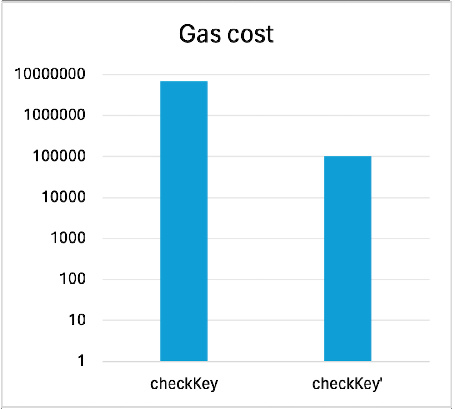}
}
\caption{\label{fig:cost_of_checkkey}Gas cost of single $\checkkey$ and $\checkkey'$}
%
\end{minipage}
\begin{minipage}[t]{0.24\textwidth}
\centering
\scalebox{0.5}{
\includegraphics{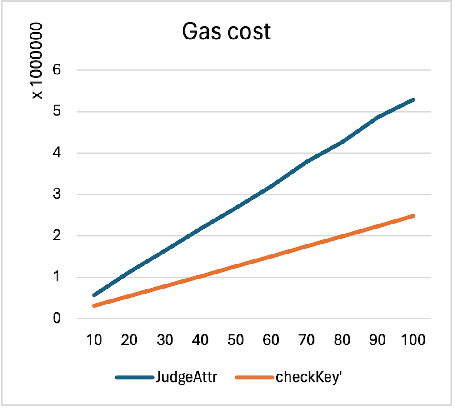}
}
\caption{\label{fig:cost_of_vrf}Total gas cost of $\judgeattrs$ and $\checkkey'$}
\end{minipage}
\begin{minipage}[t]{0.24\textwidth}
\centering
\scalebox{0.5}{
\includegraphics{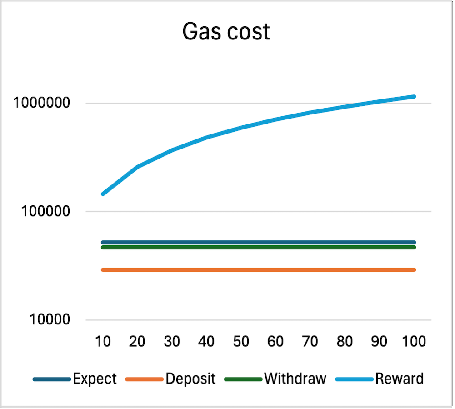}
}
\caption{\label{fig:cost_of_reward}Gas cost of incentive smart contract}
\end{minipage}
\end{figure*}

\begin{figure*}[!b]
\begin{minipage}[t]{0.32\textwidth}
\centering
\scalebox{0.65}{
\begin{tikzpicture}
\begin{axis}[
  xlabel=Number of attributes in $acp$,
  ylabel=,
  legend pos=north west,
  ymajorgrids=true,
  grid style=dashed,
]
\addplot [color=blue,mark=o]
  coordinates {(10, 15) (20, 30) (30, 44) (40, 59) (50, 74) (60, 89) (70, 103) (80, 118) (90, 133) (100, 147)};    
\addplot [color=red,mark=x]
  coordinates {(10, 155) (20,155) (30,155) (40,155) (50,155) (60,155) (70,155) (80,155) (90, 155) (100,155)}; 
\addplot [color=white]
  coordinates {(10, 255)};   
\legend{CP-ABE ciphertext size (kByte), CP-ABE key size (Byte)}
\end{axis}
\end{tikzpicture}
}
\caption{\label{fig:cpabe_size}Cipheretxt size (KByte) and key size (Byte)}
\end{minipage}
\begin{minipage}[t]{0.32\textwidth}
\scalebox{0.65}{
    \begin{tikzpicture}
        \begin{axis}[
            view={-60}{30}, 
            xlabel={\rotatebox{40}{Plaintext size (MB)}},
            ylabel={\rotatebox{-10}{Number of attributes in $acp$}},
            zlabel={Time cost (ms)},
            grid=major,
            point meta min=0,
            point meta max=1,
            ylabel style={at={(axis description cs:0.6,-0.05)}, anchor=east}, xlabel style={at={(axis description cs:1.1, 0.02)}, anchor=east}
        ]
        \addplot3[
            surf,
        ]
        coordinates {
        (10, 10, 255) (10, 20, 376) (10, 30, 480) (10, 40, 589) (10, 50, 705) (10, 60, 798) (10, 70, 875) (10, 80, 980) (10, 90, 1164) (10, 100, 1283)
        
        (20, 10, 305) (20, 20, 453) (20, 30, 557) (20, 40, 646) (20, 50, 754) (20, 60, 857) (20, 70, 947) (20, 80, 1104) (20, 90, 1251) (20, 100, 1344)
        
        (30, 10, 351) (30, 20, 477) (30, 30, 598) (30, 40, 708) (30, 50, 824) (30, 60, 936) (30, 70, 994) (30, 80, 1147) (30, 90, 1283) (30, 100, 1373)
        
        (40, 10, 419) (40, 20, 522) (40, 30, 619) (40, 40, 769) (40, 50, 887) (40, 60, 986) (40, 70, 1075) (40, 80, 1156) (40, 90, 1311) (40, 100, 1400)
        
        (50, 10, 481) (50, 20, 577) (50, 30, 680) (50, 40, 826) (50, 50, 902) (50, 60, 1015) (50, 70, 1099) (50, 80, 1181) (50, 90, 1334) (50, 100, 1462)
        
        (60, 10, 521) (60, 20, 652) (60, 30, 740) (60, 40, 851) (60, 50, 958) (60, 60, 1092) (60, 70, 1121) (60, 80, 1251) (60, 90, 1357) (60, 100, 1471)
        
        (70, 10, 575) (70, 20, 691) (70, 30, 814) (70, 40, 917) (70, 50, 999) (70, 60, 1153) (70, 70, 1213) (70, 80, 1286) (70, 90, 1479) (70, 100, 1560)
        
        (80, 10, 607) (80, 20, 730) (80, 30, 876) (80, 40, 979) (80, 50, 1116) (80, 60, 1208) (80, 70, 1291) (80, 80, 1427) (80, 90, 1481) (80, 100, 1581)
        
        (90, 10, 686) (90, 20, 827) (90, 30, 914) (90, 40, 1012) (90, 50, 1172) (90, 60, 1245) (90, 70, 1367) (90, 80, 1516) (90, 90, 1603) (90, 100, 1627)
        
        (100, 10, 704) (100, 20, 827) (100, 30, 945) (100, 40, 1052) (100, 50, 1180) (100, 60, 1270) (100, 70, 1450) (100, 80, 1607) (100, 90, 1713) (100, 100, 1799)

        };
        \end{axis}
    \end{tikzpicture}}
    \caption{Encryption cost (ms)}\label{fig:encryption_cost}    
\end{minipage}
\begin{minipage}[t]{0.32\textwidth}
\scalebox{0.65}{
\begin{tikzpicture}
    \begin{axis}[
        view={-60}{30}, 
        xlabel={\rotatebox{40}{Ciphertext size (MB)}},
        ylabel={\rotatebox{-10}{Number of attributes in $acp$}},
        zlabel={Time cost (ms)},
        grid=major,
        point meta min=0,
        point meta max=1,
        ylabel style={at={(axis description cs:0.6,-0.05)}, anchor=east}, xlabel style={at={(axis description cs:1.1, 0.02)}, anchor=east}
    ]
    \addplot3[
        surf,
    ]
    coordinates {
        (10, 10, 248) (10, 20, 352) (10, 30, 452) (10, 40, 576) (10, 50, 688) (10, 60, 775) (10, 70, 896) (10, 80, 1048) (10, 90, 1131) (10, 100, 1245)
                
        (20, 10, 321) (20, 20, 404) (20, 30, 502) (20, 40, 615) (20, 50, 708) (20, 60, 839) (20, 70, 912) (20, 80, 1057) (20, 90, 1150) (20, 100, 1261)
        
        (30, 10, 367) (30, 20, 498) (30, 30, 597) (30, 40, 702) (30, 50, 804) (30, 60, 916) (30, 70, 1046) (30, 80, 1116) (30, 90, 1221) (30, 100, 1290)
        
        (40, 10, 432) (40, 20, 523) (40, 30, 630) (40, 40, 747) (40, 50, 868) (40, 60, 1004) (40, 70, 1074) (40, 80, 1179) (40, 90, 1328) (40, 100, 1409)
        
        (50, 10, 494) (50, 20, 589) (50, 30, 698) (50, 40, 814) (50, 50, 942) (50, 60, 1042) (50, 70, 1164) (50, 80, 1310) (50, 90, 1374) (50, 100, 1479)
        
        (60, 10, 554) (60, 20, 658) (60, 30, 743) (60, 40, 869) (60, 50, 974) (60, 60, 1083) (60, 70, 1194) (60, 80, 1302) (60, 90, 1453) (60, 100, 1552)
        
        (70, 10, 605) (70, 20, 686) (70, 30, 791) (70, 40, 905) (70, 50, 1016) (70, 60, 1114) (70, 70, 1257) (70, 80, 1401) (70, 90, 1480) (70, 100, 1636)
        
        (80, 10, 689) (80, 20, 788) (80, 30, 892) (80, 40, 1004) (80, 50, 1148) (80, 60, 1242) (80, 70, 1376) (80, 80, 1505) (80, 90, 1634) (80, 100, 1731)
        
        (90, 10, 742) (90, 20, 805) (90, 30, 907) (90, 40, 1012) (90, 50, 1175) (90, 60, 1295) (90, 70, 1406) (90, 80, 1546) (90, 90, 1705) (90, 100, 1754)
        
        (100, 10, 758) (100, 20, 856) (100, 30, 973) (100, 40, 1075) (100, 50, 1225) (100, 60, 1354) (100, 70, 1436) (100, 80, 1604) (100, 90, 1738) (100, 100, 1768)
        
    };
    \end{axis}
\end{tikzpicture}}
\caption{Decryption cost (ms)}\label{fig:decryption_cost}    
\end{minipage}
\end{figure*}

In the proposed framework, data owners use $\abeencrypt$ to protect the privacy of their data; authorities invoke $\abekeygenenc$ to generate CP-ABE decryption keys; blockchain audit the authorities' honesty by invoking $\checkkey$ and $\judgeattrs$; data users apply $\getkey$ and $\abedecrypt$ to get the shared data of data owners.

We have replace the $\abekeygen$ with $\abekeygenenc$ to protect the privacy of CP-ABE decryption keys. Figure~\ref{fig:key_time_cost} shows the time cost of off-chain operations for each attribute. It shows that both $\abekeygen$ and $\abekeygenenc$ algorithms cost about 1.3ms on curve bn128. The corresponding NIZK proofs generated by the $\genProofs$ algorithm cost about 1.4ms. The $\getkey$ algorithm, which calculates the original decryption key, costs about 0.19ms. 

Due to high cost of $\GG_2$ operation in Ethereum\footnote{\url{https://github.com/musalbas/solidity-BN256G2}, Accessed: 2024-05-18}, $\checkkey$ defined in Section~\ref{sec:accountability} is slightly modified. We use native-supported $\GG_1$ operation and apply pairing-check to replace the $\GG_2$ operation to lower the gas consumption.
\begin{align}
&\hspace{-1cm}\underline{\checkkey'(\ek_0,\ek_1,\proofs,\gid,u):}\notag\\ 
\hspace{1cm}&(\pku)^{w_1} H(\gid)^{w_2}F(u)^{w_3}\overset{\text{?}}{=}\ek_0^\prime\cdot \ek_0^c\notag\\
&g_1^{w_3}\overset{\text{?}}{=}\ek_2^\prime\cdot \ek_2^c\notag\\
&e(\ek_2, g_2)\overset{\text{?}}{=}e(g_1, \ek_1)\notag\\
&e(\pku, g_2^{\alpha_\theta})e(H(\gid),g_2^{\beta_\theta})e(F(u),\ek_1) \overset{\text{?}}{=}e(\ek_0,g_2) \notag
\end{align}
where $\ek_2=g_1^{d_\theta},\ek_2^\prime=g_1^{d_\theta^\prime}$ and they are appended into NIZK $\proofs$.

Figure~\ref{fig:cost_of_checkkey} shows the gas cost of judging whether a decryption key is valid. 
The gas consumption for each attribute using $\checkkey$ and $\checkkey'$ are about $6.8\times 10^6$ and $0.1 \times 10^6$, respectively. Apparently, $\checkkey'$ is about much more gas-efficient than $\checkkey$, as predicted.

\renewcommand{\arraystretch}{1.22}
\begin{table*}[h]
\centering
{
    \caption{\centering Comparison regarding key accountability. The notations $n$, $P$ and $E$ denote the number of attributes, bilinear mapping cost and exponentiation cost, respectively}\label{table:comparison_accountability}
    \centering
    \scalebox{0.92}{
    \begin{tabular}{ |c|c|c|c|c|c|c| }
        \hline 
         Reference & Key protection  & Key generation & Proof generation& Key size &Proof size & Verification\\\hline 
        Han et al.~\cite{Han2020CPABE} & \xmark & $(2n+5)E$ & - & $(n+4)|\GG|+(logn+1)|\ZZ_p|$ & - & $(3n+4)P+(n+2)E$ 
        \\
        Yang et al.~\cite{Yang2022CPABE} & \xmark & $5nE$ & - & $3n|\GG|$+ $2|\ZZ_p|$ & - &$5nP + 4nE $
        \\
        Guo et al.~\cite{Guo2023DS} & \xmark & $(4n+4)E$ & - &  $(2n+3)|\GG|$+ $|\ZZ_p|$ & - & $(4n+4)P + (2n+2)E $
        \\
        \textbf{Ours} & \cmark & $(2n+2)E$ & $4nE$ &  $2n|\GG|$ & $2n|\GG|$+ $(3n+1)|\ZZ_p|$ & $(2n+2)P+6nE$ 
        \\\hline 
    \end{tabular}	
    }
}
\end{table*}

Smart contract uses $\judgeattrs$ to check whether a combination of keys comprises a valid decryption key for specific CP-ABE ciphertexts. 
By randomly choosing AND and OR gates in the access control policy, we evaluate the gas cost of $\judgeattrs$ and $\checkkey'$ for all attributes in Figure~\ref{fig:cost_of_vrf}. The total validation cost, as $\vrf$ needs, can be calculated by directly adding the gas cost of $\judgeattrs$ and $\checkkey'$. 

Figure~\ref{fig:cost_of_reward} depicts the gas cost of functionalities in the incentive smart contract. It can be indicated that the gas consumption for $\expect$, $\deposit$ and $\withdraw$ is constant. The gas consumption are fixed at 52248, 29182 and 47035, respectively. The gas cost of $\reward$ functionality increases linearly with the number of honest authorities involved in a data sharing process.

Figure~\ref{fig:cpabe_size} demonstrates the CP-ABE ciphertext size and (encrypted) key size. It can be seen that the ciphertext size is about 147KB when 100 attributes appeared in $acp$ and the key size is fixed at 155B. Each attribute authority costs a constant time, i.e., 1.23ms, to generate a key for each attribute string.
We have combined CP-ABE with symmetric encryption, i.e., leveraging CP-ABE plaintext as symmetric key seed. Since KDF is employed to generate symmetric key and XOR is leveraged as encryption/decryption operation. Further, we evaluate the performance of these offchain operations using Golang implementation.  
Figure~\ref{fig:encryption_cost} and Figure~\ref{fig:decryption_cost} depict the total encryption and decryption cost considering both attributes number and plaintext/ciphertext size, respectively. With 100 attributes in $acp$ and 100MB plaintext/ciphertext, the total time overhead of encryption (or decryption) is about 1.8s.

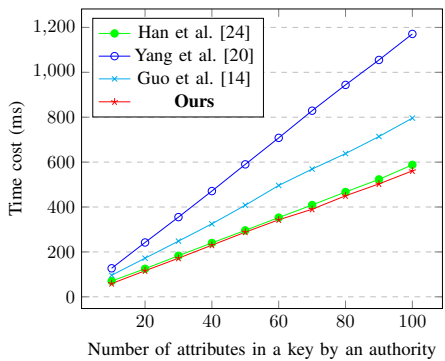
\begin{figure}[!h]
\begin{minipage}[t]{0.45\textwidth}
\centering
\scalebox{0.7}{
\begin{tikzpicture}
\begin{axis}[
  xlabel=Number of attributes in a key by an authority,
  ylabel=Time cost (ms),
  legend pos=north west,
  ymajorgrids=true,
  grid style=dashed,
]

\addplot [color=green,mark=*]
  coordinates {(10, 70) (20, 125) (30, 183) (40, 240) (50,296) (60,353) (70, 409) (80,467) (90, 523) (100,588)};
\addplot [color=blue,mark=o]
  coordinates {((10, 127) (20, 242) (30, 355) (40, 471) (50, 590) (60, 708) (70, 829) (80, 944) (90, 1055) (100, 1171)};
\addplot [color=cyan,mark=x]
  coordinates {(10, 96) (20, 172) (30,248) (40, 325) (50, 408) (60, 496) (70, 569) (80, 638) (90, 714) (100, 796)};
\addplot [color=red,mark=star]
  coordinates {(10, 59) (20, 116) (30,172) (40, 230) (50, 288) (60, 343) (70, 390) (80, 450) (90, 503) (100, 561)};
\legend{Han et al.~\cite{Han2020CPABE}, Yang et al.~\cite{Yang2022CPABE}, Guo et al.~\cite{Guo2023DS}, \textbf{Ours}}
\end{axis}
\end{tikzpicture}
}
\caption{\label{fig:verification_comparison}The computational overhead regarding key verification or accountability}
\end{minipage}
\end{figure}

Denote $E$ as an exponentiation on a group and $P$ as a bilinear pairing. From Table~\ref{table:group_op_cost}, we can know that pairing is much more costing than exponentiation. Han et al.~\cite{Han2020CPABE}, Guo et al.~\cite{Guo2023DS} and Yang et al.~\cite{Yang2022CPABE} are some of the most recent decentralized CP-ABE with accountability. Table~\ref{table:comparison_accountability} compares the performance regarding CP-ABE key generation, the associating NIZK proof generation and key accountability/verification. It can be seen that only our protocol protects key privacy and other works~\cite{Han2020CPABE,Guo2023DS,Yang2022CPABE} expose keys when considering key accountability. As a sequence, we encounter additional overhead to generate the NIZK proofs, which is verified by the $\checkkey$ algorithm. In $\checkkey$, $\pk_u$, $g_2^{\alpha_\theta}$, $H(\gid)$ and $g_2^{\beta_\theta}$ are public available. Hence, only $2n+2$ bilinear pairings is required, where $n$ is the number of attributes appeared in a key from a single authority. Figure~\ref{fig:verification_comparison} indicates that our protocol requires lower computational overhead in key verification (or accountability) than other works~\cite{Han2020CPABE,Guo2023DS,Yang2022CPABE}.

\subsection{Property analysis}
\label{sec:th_analysis}

The proposed scheme satisfies the system goals defined in Section~\ref{sec:goals} and we make a analysis as follows.
\begin{enumerate}
    \item \textit{Tamper-resistance}: Tamper-resistance of the proposed scheme directly inherits from the property of blockchain. With Ethereum blockchain storing encrypted CP-ABE keys, keys are untamperable. Also, \textbf{DSP} is a buisiness-related trusted service. Thus, encrypted messages or files are unalterable. 
    \item \textit{Decentralization}: On the one hand, Ethereum blockchain is a fully decentralized system, without a privileged administrator or manager. On the other hand, decentralized CP-ABE is leveraged as the underlying cryptographic primitive, allowing multiple attribute authorities to make a decision collectively in data sharing.
    \item \textit{Key privacy}: We slightly modify the key generation algorithm so that the output protects privacy of a valid CP-ABE key, as $\abekeygenenc$ shows. We also demonstrate that the algorithm is resistant to eavesdropping attacks by Lemma~\ref{lem:enckey_secure}. The targeted user can recover the legal decryption key with its private key, as $\getkey$ demonstrates. Besides, we argue that the malleability key described in Lemma~\ref{lem:2} cannot generate decryption keys of new attributes. Thus, the output of $\abekeygenenc$ can be delivered publicly through blockchain without leaking key privacy.
    \item \textit{Data privacy}: \textbf{DSP} provides available data storage to authorized users. To resist honest but curious \textbf{DSP}, data user encrypts its data by applying RW CP-ABE before uploading the data. The underlying RW CP-ABE is secure based on the q-DPBDHE2 assumption~\cite{Rouselakis2015EfficientSL}. We have proved that the public output of $\abekeygenenc$ does not reveal information about $(K_0, K_1)$. We leverage $\judgeattrs$ to judge whether a user is authorized to access ciphertext stored on \textbf{DSP}. The judgment is accomplished publicly on Ethereum smart contract. Lemma~\ref{lem:4} and Lemma~\ref{lem:5} show that owner's ciphertext and digital asset are private against malicious users. Thus, only qualified users (with legal keys and enough deposits) can access the ciphertext, further eliminating the risks of data leakage. 
    \item \textit{Forward-secrecy}: $\gid$ enables the decentralized CP-ABE scheme to resist collusion attack~\cite{Rouselakis2015EfficientSL,Lewko2011DABE}, making CP-ABE keys unaggregatable given different $\gid$. In our design, each data-sharing instance corresponds to a unique $\gid$, making different sharing instances separately. Thus, a decryption key is only valid for a unique ciphertext. Moreover, even if a malicious user gets a valid decryption key, it cannot access the ciphertext from \textbf{DSP}, as clarified by Lemma~\ref{lem:4}. Hence, when a decryption key is leaked, the privacy of previously shared data is still guaranteed.
    \item \textit{Incentives}: In our protocol, we design an owner-driven approach --- data owner sets the expected benefit when sharing its data using $\expect$ on smart contract. Subtracting the required digital currency of data owner, the rest of deposited currency is transferred to honest authorities as rewards. Hence, both data owner and honest authorities will gain incentives in the paradigm.
    \item \textit{Accountability}: The proposed framework require authorities to issue CP-ABE keys using the $\abekeygenenc$ algorithm. Additionally, NIZK proofs are attached for the encrypted CP-ABE keys, as demonstrated by the $\genProofs$ algorithm. Lemma~\ref{lem:3} proves that authorities are accountable with the NIZK proofs. Moreover, the accountability is automatically accomplished on smart contract.
    \item \textit{Key abuse resistance}: We first consider the illegal authorization problem. The \textit{Accountability} property guarantees that authorities cannot issue invalid keys. Besides, the process of key issuance (i.e., authorization) to each target user is transparent and publicly verifiable. Then, we discuss the key disclosure problem.
    Lemma~\ref{lem:enckey_secure} and Lemma~\ref{lem:2} highlight that malicious users cannot construct meaningful keys. Lemma~\ref{lem:5} implies that exposing key to another user is useless in obtaining data owner's digital assets. Hence, key disclosure problem is resolved.
\end{enumerate}

\section{Related works}
\subsection{Accountability in CP-ABE}
Liu et al.~\cite{Liu2012TCPABE} propose to reduce trust in authorities by introducing a central authority (CA) and tracing authority (TA) in a white-box traceable CP-ABE scheme. 
Later, Yu et al.~\cite{Yu2017ACPABE} put the user's identity into the CP-ABE decryption key. Thus, any third party can recognize the user directly from the exposed decryption key. They use a digital signature to make authorities publicly verifiable. To uncover the misbehavior of malicious authorities, they put forward two kinds of authorities: namely, central authority (CA) and outsourced authority (OA). These two authorities are assumed not to collude with each other. 

Ning et al.~\cite{Ning2015ACPABE} define an accountable authority for the first time. If a malicious authority distributes legitimate keys to unauthorized users, such behaviors can be discovered. The main idea is to make the data user's decryption key jointly determined by the authority and the user. 
Later, Li et al.~\cite{Li2020HCPABE} compose a CP-ABE scheme achieving authority accountability in their proposed hidden-policy ABE scheme. 
More recently, Liu et al.~\cite{Liu2022BACPABE} put forward a black-box accountable CP-ABE scheme to identify the owner of the faked decryption device and the malicious authority. 
Hei et al.~\cite{Hei2021ABE} employ a dual receiver cryptosystem (DRC)~\cite{Diament2011DRC} to generate key pairs for users and the key auditor. In the proposed CP-ABE algorithm, the authorities generate and encrypt keys using the corresponding user's public key.
Guo et al.~\cite{Guo2023DS} uses CP-ABE to protect privacy in VANET data sharing, where a consortium blockchain is incorporated to resist key abuse of AAs. For user key abuse, they achieve white-box traceability, as previous researches~\cite{Liu2012TCPABE,Ning2015ACPABE,Li2020HCPABE} do.

\subsection{Key abuse in CP-ABE}

Chase et al.~\cite{Chase2009ABESP} introduce an anonymous key issuing protocol based on a two-party secure computation (2PC) protocol to enhance user privacy. 
Qiao et al.~\cite{Qiao2018CPABE} regard the key disclosure problem as privilege abuse. Then, they present a black-box compulsory traceable CP-ABE scheme by introducing a tracing ciphertext. The scheme guarantees that the adversary cannot distinguish the tracing ciphertext from the normal ciphertext, thus addressing privilege abuse.
Cui et al.~\cite{Cui2020RFCPABE} define the concept of key regeneration-free for CP-ABE, aiming to eliminate key abuse by illegal data users. The key idea is to introduce randomness in different parts of the decryption key. Later, Zhang et al.~\cite{Zhang2021RKACPABE} apply the key regeneration-free technique to achieve strong key unforgeability in a lightweight data sharing scheme. Hei et al.~\cite{Hei2021ABE} design a key audit scheme using an incentive mechanism to address the key leakage issue. To protect the privacy of attribute keys, they incorporate dual receiver cryptosystem techniques. Meanwhile, the illegal authorization issue is addressed through transparent storage on the blockchain. However, this mechanism relies on a trusted role (auditor), who is used to detect malicious behaviour. 

Recently, Liu et al.~\cite{Liu2022KM} conduct comprehensive key management in CP-ABE for cloud data with a consortium blockchain. Specifically, they leverage Hyperledger Fabric~\cite{Androulaki2018Fabric}, where distributed nodes collectively keep secret keys via Pedersen's distributed key generation (DKG) technique. When using DKG, they leverage blockchain to achieve consensus among validators.
They further divide decryption phase into two parts. Partial decryption uses a publicly known decryption key from the blockchain and full decryption makes use of a user secret key. Consequently, their approach achieves key escrow-free and key distribution-free. Moreover, they are able to trace the illegal user identity without exposing the CP-ABE decryption key. However, they do not provide the feature of authority accountability.

\renewcommand{\arraystretch}{1.3}

\begin{table*}[!t]
    \caption{Comparison with related works}\label{table:protocol_comparison}
    \centering
    \scalebox{1}{
    \begin{tabular}{@{}lcccccccc@{}}
        \hline
        \hline
        {\bfseries Approach} &  
        {\bfseries Data sharing}  & 
        {\bfseries Decentralization} & 
        {\bfseries Accountability} & 
        {\bfseries Forward-secrecy} & 
        {\bfseries Key-privacy}  & 
        {\bfseries Incentives} & 
        {\bfseries Key abuse resistance}  \\
        \hline
        \hline        
        Sifah et al.~\cite{Sifah2021SelectiveS} & \cmark & \cmark & \cmark & \xmark & \xmark & \xmark &  \xmark \\
        Li et al.~\cite{Li2020HCPABE} & \cmark & \xmark & \cmark & \xmark & \xmark & \xmark  &  \cmark \\
        Chase et al.~\cite{Chase2009ABESP} & \xmark & \cmark & \xmark & \cmark & \xmark & \xmark  & \cmark \\
        Liu et al.~\cite{Liu2022KM} & \cmark & \cmark & \xmark & \cmark & \xmark & \xmark  &  \cmark \\
        Zhang et al.~\cite{Zhang2023DS} & \cmark & \xmark & \cmark & \xmark & \xmark & \xmark &  \xmark \\
        Guo et al.~\cite{Guo2023DS} & \cmark & \cmark & \cmark & \cmark & \xmark & \xmark  &  \cmark \\
        Hei et al.~\cite{Hei2021ABE} & \cmark & \cmark & \cmark & \xmark & \xmark & \cmark  & \cmark \\
        Ren et al.~\cite{Ren2024DS} & \cmark & \cmark & \cmark & \xmark & \cmark & \cmark  & \xmark \\
        \textbf{Ours} & \cmark & \cmark & \cmark & \cmark & \cmark & \cmark & \cmark \\
        \hline       
    \end{tabular}
    }
\raggedright
\end{table*}
\subsection{Data sharing using CP-ABE}

Ren et al.~\cite{Ren2024DS} introduce a blockchain-based CP-ABE, where ciphertext and keys are distributed and managed in blockchain. They incorporate a threshold proxy re-encryption to protect data owner's master key. They also consider a staking economic incentive model with reward or slashing to ensure nodes security. 
Qin et al.~\cite{Qin2021AC} put forward a blockchain-based multi-authority access control scheme for sharing data securely. They exploit smart contracts to compute tokens for attributes managed across multiple management domains. Moreover, blockchain is applied to record the access process so that the access history can be audited.


Wang et al.~\cite{Wang2021MedShare} develops a privacy-preserving medical data sharing system by using blockchain, thereby eliminating the single-point-of-failure problem. They design a non-interactive on-chain search protocol to enable multiple users to search and access electronic health records. Murad et al.~\cite{Murad2024IoT} propose a P2P architecture within fog
nodes to handle client requests, thereby reducing the cloud overhead and enhancing efficiency. They leverage a CP-ABE scheme to obtain a fine-grained and secure communication channel.
Mahdavi et al.~\cite{Mahdavi2024IoT} downgrade standard ABE to fuzzy identity-based encryption to lower computation in IoT devices. Further, they outsource exponentiation and bilinear pairing of CP-ABE scheme to cloud servers. Zhang et al.~\cite{Zhang2024CPABE} design a revocable CP-ABE scheme to resist economic denial of sustainability attack in data sharing, where blockchain is employed to judge the download permission of data users. 
Yang et al.~\cite{Yang2022CPABE} propose a traceable privacy-preserving data sharing for fog-based smart logistics based on multi-authority CP-ABE. 

Zhang et al.~\cite{Zhang2023DS} protect the privacy of Metaverse healthcare data by presenting a CP-ABE scheme with constant computation overhead, which applies a multi-server architecture. To minimize storage space caused by invalid and duplicated ciphertexts, they introduce a mechanism to check the validity and equivalence of ciphertexts. Additionally, they introduce an attribute-based re-encryption technique to enable authority delegation after deduplication.

Other works~\cite{Ma2020ABE, Zhang2021ABE, Xue2023ABE, Zhang2023ABE} also address data sharing using blockchain and CP-ABE, focusing on various aspects such as attribute revocation, ciphertext access strategies, update capabilities, and invisibility.

\subsection{Comparison with related works}
\label{sec:comparison}
Blockchain is a promising infrastructure for sharing data among distributed enterprises or entities.
Sifah et al.~\cite{Sifah2021SelectiveS} propose a method to selectively share outsourced encrypted data in the cloud by embedding an authorization policy to data. They claim to achieve data traceability or auditability, because blockchain is employed to record tamper-proof logs and activities. 

Liu et al.~\cite{Liu2022KM} and Guo et al.~\cite{Guo2023DS} resolve the key leakage problem by invalidating the user's key and invoking a key update algorithm, respectively.
Some systems~\cite{Chase2009ABESP,Liu2022KM,Guo2023DS} have considered the single point failure problem, hence achieving decentralization. 
Zhang et al.~\cite{Zhang2023DS} mainly focus on CP-ABE ciphertext management, rather than keys, in metaverse healthcare. Ren et al.~\cite{Ren2024DS} manage CP-ABE keys with a threshold proxy re-encryption scheme.

Table~\ref{table:protocol_comparison} summarizes the comparison with related works regarding the achieved features. Some works~\cite{Li2020HCPABE,Zhang2023DS} are not applicable to the metaverse environment due to the use of single authority CP-ABE. The 2PC protocol which enables user anonymity in Chase et al.~\cite{Chase2009ABESP} is interactive and requires a secure communication channel, leading to obstacles in application in metaverse. Sifah et al.~\cite{Sifah2021SelectiveS} focus on access control policy which enables to outsource data selectively and omit the discussion of key generation and delivery, thus inapplicable for metaverse. Other works~\cite{Hei2021ABE,Li2020HCPABE,Liu2022KM,Guo2023DS} implement traceability of CP-ABE keys, and the ``trace'' methods input CP-ABE keys. That means decryption keys should be presented for accountability, which is impractical in a public metaverse environment. The sharing scheme provided by Ren et al.~\cite{Ren2024DS} can be applied in metaverse with a different system model, since additional proxies should be considered.

\section{Application in GameFi}
GameFi, allowing players to earn real-world rewards through gameplay, serves as a financial layer that enriches the interactive and economic aspects of the metaverse. GameFi often incorporates elements like play-to-earn (P2E) mechanics, governance tokens, and decentralized autonomous organizations (DAOs), enabling players not only to earn income but also to participate in decision-making processes that shape the game's development and broader ecosystem. By merging gaming and financial incentives, GameFi aims to foster a more engaging and community-driven metaverse.

If the game provider (i.e., \textbf{DSP} in this context) establishes all the rules, it may reduce player engagement and trust in the game. Such centralized control can lead players to feel a lack of autonomy and fairness, which can diminish their motivation to participate. 
To enhance the game's appeal, providers should consider adopting more open and transparent rules, encouraging player involvement in the decision-making process, and fostering a greater sense of belonging and investment among players.
However, ensuring the security and privacy of user data and digital assets is paramount in GameFi.

The proposed decentralized CP-ABE based framework with accountable authority offers a compelling solution to address these challenges. 
With the framework, players can stake their money in smart contract to become CP-ABE authorities and they can interact with game provider with special APIs. Then, each user attribute (e.g., age, level, role, achievement, region, sex, playtime, skill, membership, balance, etc.) is accessible only to specific authorities.  
Essentially, these authorities form a decentralized autonomous organization (DAO)~\cite{Gadekallu2023Meta} with accountability.
Game players or providers can leverage the CP-ABE encryption algorithm to secure various virtual assets, including skins, weapons, and other in-game items. By defining access policies that correspond to user attributes, players or providers can ensure that only those who meet the criteria can access or trade these assets. 
This not only enhances security but also creates a more engaging gaming experience, as players are rewarded based on their accomplishments. 

Suppose each in-game item is bound to an NFT~\cite{EntrikenERC721} address. When a player (or the game provider) wishes to trade an in-game item, he/she encrypts the description, which demonstrates the NFT ownership, with a proper access control policy. 
For example, a skin of player1 in an online game is encrypted with an access control policy $acp$=((level$\geq$25@AUTH1 OR cityLA@AUTH2) AND female@AUTH3). Then, player2 with attribute $[u_1]$=[level25@AUTH1, cityPHX@AUTH2, female@AUTH3] can request to authorities for decryption keys to obtain the ownership of the skin. However, player3 with attributes $[u_2]$=[level28@AUTH1, cityLA@AUTH2, male@AUTH3] is unable to buy the item. AUTH1, AUTH2 and AUTH3 are different authority identities, demonstrating that each authority has separated attribute spaces. As described by the framework, the encrypted keys sent to player2 are generated using $\abekeygenenc$ algorithm, along with $\genProofs$ being invoked. The encrypted keys and proofs are uploaded to smart contract. Finally, player2 can run $\getkey$ to obtain valid keys and execute the CP-ABE decryption algorithm to obtain player1's on-sale skin. After a successful trade of an in-game item, the authorities gain rewards for providing key issuance services, as the $\reward$ algorithm shows. In case an authority is caught misbehaving in the service which can be detected by the $\checkkey$ algorithm, its stake will be forfeited by the smart contract.

Similarly, the game provider, acting as a data seller, can issue new in-game merchandise through this framework. This approach minimizes the risks associated with the arbitrary issuance of merchandise by the game provider, enhancing transparency and credibility, as the entire inventory of merchandise is publicly accessible.

Clearly, the proposed framework will undoubtedly increase players' enthusiasm for participating in the game, thereby enhancing the overall prosperity of the game ecosystem.


\section{Conclusion and future work}
This paper proposes a blockchain-driven data sharing paradigm for the metaverse based on the RW decentralized CP-ABE scheme. The data sharing paradigm achieves secure key management, where the key abuse problem is addressed and authority accountability is considered. We leverage blockchain to store encrypted CP-ABE keys, to which NIZK proofs are attached. The keys are verified publicly in smart contract after authorities publish them. Besides, data users are qualified to obtain the keys only if they deposit enough digital currency in another smart contract. Data owners and authorities are rewarded after users access data successfully. Hence, the proposed paradigm gets rid of the key disclosure and illegal authorization problems. Compared with related works, we are the first to realize authority accountability with key privacy. 

We plan to further refine the application of the proposed framework in GameFi, such as enabling DAO governance and implementing in-game item rentals and exchange.
Additionally, blockchains and smart contracts inherently lack the ability to access data outside their network (i.e., off-chain data). In the future, we plan to build a decentralized oracle based on the proposed data sharing paradigm, making full use of accountability of CP-ABE authorities and fine-grained access control. The oracle aims at providing services that manage external data outside blockchain in a trustworthy and decentralized manner.


\begin{IEEEbiography}[{\includegraphics[width=1in,height=1.25in,clip,keepaspectratio]{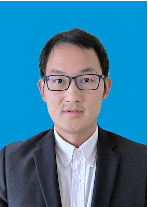}}]{Liang Zhang} obtained excellent programming skills in Baidu Co., Ltd. after acquiring bachelor's degree from Huazhong University of Science and Technology. He then acquired Ph.D. degree from Fudan University in 2022. He is currently an associate professor in Hainan University and a postdoctoral fellow at Hong Kong University of Science and Technology. His research interests include blockchain and applied cryptography. His recent publications are included in \emph{IEEE TIFS}, \emph{IEEE TSC}, \emph{CJ}, \emph{CF'22} and \emph{SACMAT'22}.
\end{IEEEbiography}

\begin{IEEEbiography}[{\includegraphics[width=1in,height=1.25in,clip,keepaspectratio]{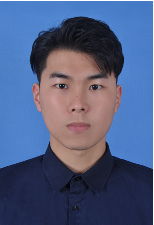}}]{Zhanrong Ou} was born in 1998. He received a bachelor's degree in Hanshan Normal University, Chaozhou, Guangdong province, in 2021. He joined Hainan University in September 2022. Now, he is pursuing his master's degree in the School of Cyberspace Security (School of Cryptology) at Hainan University. His research interests consist of blockchain and decentralized randomness beacon.
\end{IEEEbiography}
 
\begin{IEEEbiography}[{\includegraphics[width=1in,height=1.25in,clip,keepaspectratio]{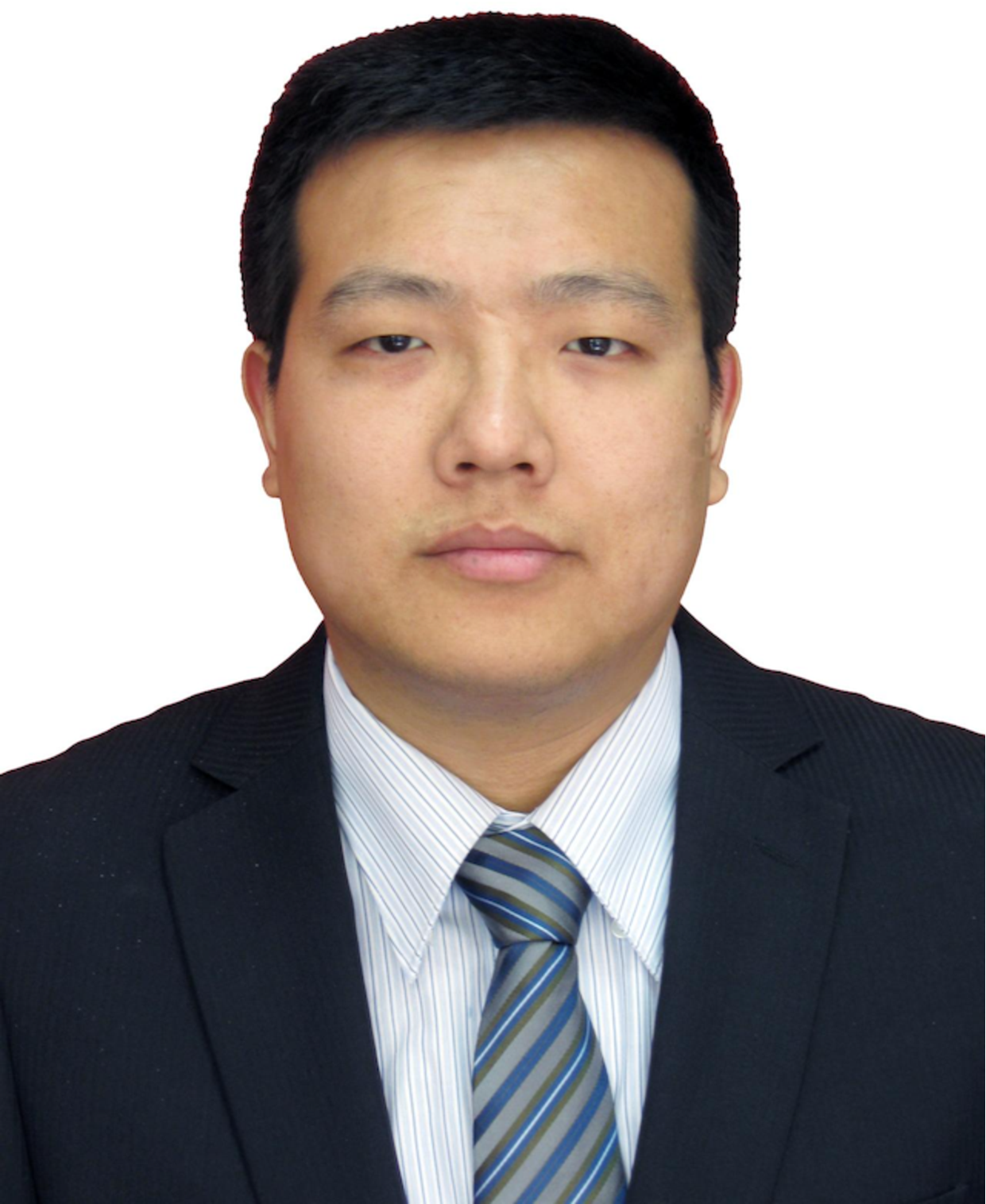}}]{Changhui Hu} received the PhD degree from the School of Mathematics, Shandong University, in 2012. He is currently working as a professor with the School of Cyberspace Security (School of Cryptology), Hainan University. His research interests include information security and cryptography. His recent publications are included in \emph{Usenix Security'21}, \emph{IEEE TCC}, \emph{JIS}, \emph{CN} and \emph{SCN}.  
\end{IEEEbiography}

\begin{IEEEbiography}[{\includegraphics[width=1in,height=1.25in,clip,keepaspectratio]{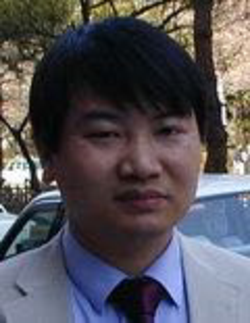}}]{Haibin Kan} was born in 1971. He received a Ph.D. from Fudan University, Shanghai, China, 1999. From June 2002 to February 2006, he was with the Japan Advanced Institute of Science and Technology as an assistant professor. He went back Fudan University in February 2006, where he is currently a full professor. He is also the Director of the Shanghai Blockchain Engineering Research Center. His research interests include coding theory, cryptography, and computation complexity. 
\end{IEEEbiography}

\begin{IEEEbiography}[{\includegraphics[width=1in,height=1.25in,clip,keepaspectratio]{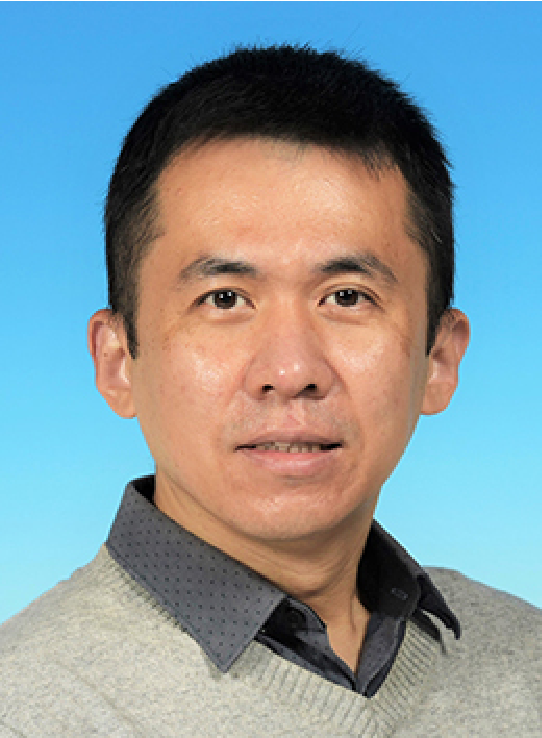}}]{Jiheng Zhang} received both Ph.D. and M.S. degrees in operations research from Georgia Institute of Technology, MS degree in mathematics from the Ohio State University, and BS degree in mathematics in Nanjing University. He joined Industrial Engineering and Decision Analytics (IEDA) of the Hong Kong University of Science and Technology (HKUST) as an Assistant Professor in 2009, and was promoted to Associate Professor in 2015 and then to full Professor in 2020. He now is the department head of IEDA of HKUST. He is also the associate director of HKUST Crypto-Fintech Lab. 
\end{IEEEbiography}

\vfill

\newpage

\appendix

\subsection{Security Properties of Sigma Protocol}
\label{sec:security_properties}

\begin{myDef}\label{def:sigma_properties} Assuming a well-randomized setup and random oracle, Sigma protocol satisfies the following security properties:
\begin{enumerate}
\item \textbf{Correctness:} If P is honest, honest V outputs $True$ with high probability.
\item \textbf{Knowledge soundness:} Given two correct conversations $(b, c, \{w_1,\cdots, w_n\})$ and $(b, c^\prime, \{w_1^\prime,\cdots, w_n^\prime\})$ where $c\neq c^\prime$, it is efficient to calculate $X$. 
\item \textbf{Special honest verifier zero knowledge:} There exists a polynomial-time simulator that can convince V without revealing any additional information. That means, the proof $\proofs$ reveals nothing information about $X$.
\end{enumerate}
\end{myDef}

\subsection{Security Proof}
\label{sec:security_proof}
We prove the required security properties of for the modified RW CP-ABE $\checkkey$ algorithms as below. For convenience, we omit footnote $\theta$. 

\subsubsection{\textbf{Correctness}}
We prove \textbf{correctness} of Equations~\eqref{eq:checkkey1},~\eqref{eq:checkkey2},~\eqref{eq:checkkey3} from left to right, respectively.
\begin{align*}
left1&=(\pku)^{w_1} H(\gid)^{w_2}F(u)^{w_3} \notag\\
&=(\pku)^{\alpha^\prime+c\alpha} H(\gid)^{\beta^\prime+c\beta}F(u)^{d^\prime+cd}\notag\\
&=(\pku)^{\alpha^\prime}H(\gid)^{\beta^\prime}F(u)^{d^\prime}\cdot ((\pku)^{\alpha} H(\gid)^{\beta}F(u)^{d} )^c\notag\\
&=\ek_0^\prime\cdot \ek_0^c\notag\\
&=right1 \notag\\
\notag\\
left2&=g_2^{w_3}=g_2^{d^\prime+c\cdot d}=g_2^{d^\prime}\cdot (g_2^{d})^c=\ek_1^\prime\cdot \ek_1^c\notag\\
&=right2\notag\\
\notag\\
left3&=e(\pku, g_2^{\alpha})e(H(\gid),g_2^{\beta})e(F(u),\ek_1) \notag\\
&=e(g_1^{y{\alpha}}, g_2)e(H(\gid)^{\beta}, g_2) e(F(u)^{d}, g_2)\notag\\
&=e(g_1^{y\alpha}H(\gid)^{\beta}F(u)^{d}, g_2)\notag\\
&=e(\ek_0,g_2)\notag\\
&=right3\notag\\
\end{align*}

\subsubsection{\textbf{Knowledge soundness}}
According to definition of \textbf{Knowledge soundness}, two accepting conversations (i.e., NIZK proofs) $\proofs$ and $\proofs'$ are given, i.e.,\\ 
$\proofs=\left\{\begin{array}{l}
    \ek_0^\prime=(\pku)^{\alpha^\prime} H(\gid)^{\beta^\prime}F(u)^{d^\prime}\tabularnewline
    \ek_1^\prime=g_2^{d^\prime}\tabularnewline
    c=Hash(\ek_0, \ek_0^\prime)\tabularnewline
    w_1=\alpha^\prime + c\cdot \alpha \tabularnewline
    w_2=\beta^\prime + c\cdot \beta \tabularnewline
    w_3=d^\prime + c\cdot d
\end{array}\right.$ 
and \\
$\proofs'=\left\{\begin{array}{l}
    \ek_0^\prime=(\pku)^{\alpha^\prime} H(\gid)^{\beta^\prime}F(u)^{d^\prime}\tabularnewline
    \ek_1^\prime=g_2^{d^\prime}\tabularnewline
    c'=Hash'(\ek_0, \ek_0^\prime)\tabularnewline
    w'_1=\alpha^\prime + c'\cdot \alpha \tabularnewline
    w'_2=\beta^\prime + c'\cdot \beta \tabularnewline
    w'_3=d^\prime + c'\cdot d
\end{array}\right.$, where $Hash'$ is a different hash function so that $c'\neq c$. Note that the two conversations share the same Sigma protocol commitment tuple $(\ek_0^\prime, \ek_1^\prime)$. Then, the witness $\alpha, \beta, d$ can be calculated as below:
$\alpha = \frac{w'_1-w_1}{c'-c}, \beta = \frac{w'_2-w_2}{c'-c}, d = \frac{w'_3-w_3}{c'-c}$. 

\subsubsection{\textbf{Special honest verifier zero knowledge}}
\label{sec:special_hvzk}
The property of \textbf{special honest verifier zero knowledge} can be proved if any adversary $\adv$ learns nothing about a valid decryption key ($K_0, K_1$) from public messages (i.e., global parameters, $\ek_0$, $\ek_1$ and $\proofs$). By Lemma~\ref{lem:enckey_secure}, we learn that $(\ek_0, \ek_1)$ reveals nothing useful information about $(K_0, K_1)$. Now we prove that the conversation $\proofs$ conveys zero knowledge. 

Let's prove that $\adv$ can simulate a conversation as the prover (i.e., RW CP-ABE authority) does. However, $\adv$ does not know the prover's private key $(\alpha,\beta)$. $\adv$ calls a simulator $\simu$ to generate a conversation. $\simu$ can calculate the conversation values in an arbitrary order.
Since anyone can simulate $\ek_1=K_1=g_2^d$, where $d\xleftarrow[]{R} \ZZ_p$, we only talk about the $\ek_0$ part. 

In the simulated world, $\simu$ chooses $w_1, w_2, w_3\xleftarrow[]{R} \ZZ_p$, $c \xleftarrow[]{R} \ZZ_p$ and calculates $(\pku)^{w_3}H(\gid)^{w_2}F(u)^{w_3}/{\ek_0^c}$ as $\ek'_0$. Then, output these values in a reverse order, i.e., ($\ek'_0, c, (w_1, w_2, w_3)$). Note that the output is always an accepting conversation by $\adv$, as required. It can be inferred that the values of the conversation are uniformly distributed in their respective spaces.

In real world Sigma protocol, $\ek'_0$ is uniformly distributed, since $\alpha', \beta', d'\xleftarrow[]{R} \ZZ_p$ are randomly chosen. $c$ is also uniformly distributed in $\ZZ_p$, guaranteed by the hash function. Obviously, ($w_1, w_2, w_3$) are also uniformly distributed, since they are calculated as $w_1=\alpha'+c\cdot \alpha, w_2=\beta'+c\cdot \beta, w_3=d'+c\cdot d$. 

It can be seen that simulated conversation and real world conversation are identical.
In summary, there exists a simulator $\simu$ to generate an accepting conversation to convince verifiers. Moreover, the simulation reveals no information about the prover's secrets/witness ($\alpha, \beta$) .

\end{document}